\documentclass[twocolumn,epjc3,envcountsect]{svjour3}

\RequirePackage[T1]{fontenc}
\usepackage[american]{babel}
\usepackage[normalem]{ulem}

\usepackage{comment}

\RequirePackage[T1]{fontenc}

\smartqed

\RequirePackage{graphicx}
\RequirePackage{mathptmx}
\RequirePackage{flushend}
\RequirePackage[numbers,sort&compress]{natbib}
\RequirePackage[colorlinks,citecolor=blue,urlcolor=blue,linkcolor=blue]{hyperref}

\usepackage{amssymb}
\usepackage{color,colortbl}
\usepackage{epstopdf}
\usepackage{graphicx}
\usepackage{tabularx}
\usepackage{siunitx} 
\usepackage{booktabs}
\usepackage{amsmath}
\usepackage{dsfont}
\usepackage{xstring}
\usepackage{subfigure}

\usepackage{microtype} 

\usepackage{arydshln}

\usepackage[capitalize]{cleveref}
\Crefname{section}{Section}{Sections}
\Crefname{figure}{Figure}{Figures}
\Crefname{table}{Table}{Tables}
\Crefname{appendix}{Appendix}{Appendices}
\Crefname{equation}{Eq.}{Eqs.}
\newcommand{\Onlinecite}[1]{%
    \IfSubStr{#1}{,}{Refs}{Ref}.~\cite{#1}%
}

\DeclareMathOperator{\tr}{\rm tr\,}

\definecolor{Gray}{gray}{0.9}

\journalname{Eur. Phys. J. C}

\begin{document}

\title{Flux tubes in QCD at finite temperature}

\author{M. Baker\thanksref{e1,addr1}
\and
P. Cea\thanksref{e2,addr2}
\and
V. Chelnokov\thanksref{e3,addr3}
\and
L. Cosmai\thanksref{e4,addr2}
\and
A. Papa\thanksref{e5,addr4,addr5}
}

\institute{Department of Physics, University of Washington, WA 98105 Seattle, USA\label{addr1}
\and
INFN - Sezione di Bari, I-70126 Bari, Italy\label{addr2}
\and
Institut f\"ur Theoretische Physik, Goethe Universit\"at, 60438 Frankfurt am Main, Germany\label{addr3}
\and
Dipartimento di Fisica, Universit\`a della Calabria, I-87036 Arcavacata di Rende, Cosenza, Italy\label{addr4}
\and
INFN - Gruppo collegato di Cosenza, I-87036 Arcavacata di Rende, Cosenza, Italy\label{addr5}
}

\thankstext{e1}{e-mail: mbaker4@uw.edu}
\thankstext{e2}{e-mail: paolo.cea@ba.infn.it}
\thankstext{e3}{e-mail: chelnokov@itp.uni-frankfurt.de}
\thankstext{e4}{e-mail: leonardo.cosmai@ba.infn.it}
\thankstext{e5}{e-mail: alessandro.papa@fis.unical.it}

\date{Received: date / Accepted: date}

\maketitle

\begin{abstract}
We present results for the chromo-electric field generated by a static quark–antiquark pair at finite temperature, in lattice QCD with 2+1 dynamical staggered fermions at physical quark masses. We investigate the evolution of the field as the temperature increases through and beyond the chiral transition. 
For all the  temperatures considered
we find clear evidence of a
chromo-magnetic current and of a longitudinal nonperturbative chromo-electric field that stays almost uniform along the flux tube.
In the high-temperature region the magnitude of the flux-tube field is determined by an effective string tension that decreases
exponentially 
as the temperature increases, while the flux-tube
width decreases according to an inverse-temperature law. Our results  suggest that beyond the chiral pseudocritical temperature the quark- antiquark system can be characterized by a {\em screened} string tension.
\end{abstract}

\section{Introduction}
With the aim of deepening the microscopic understanding of color confinement, in recent years we have undertaken an extended program of numerical studies based on Monte Carlo simulations of the SU(3) Yang--Mills theory and of full QCD with (2+1) Highly Improved Staggered (HISQ) flavors, discretized on a Euclidean space--time lattice. Within this framework, we have systematically analyzed the local structure of the color fields generated by two static color sources, a quark and an antiquark, both at zero temperature~\cite{Baker:2018mhw,Baker:2019gsi,Baker:2022cwb,Baker:2024peg} and in the thermal regime~\cite{Baker:2023dnn}. These investigations have also been considered in the context of phenomenological approaches to hadronization, where a realistic modeling of confinement dynamics is essential (see, {\it e.g.}, Refs.~\cite{Bierlich:2022oja,Bierlich:2020naj}).

From this body of work, the following key results have emerged:
\begin{itemize}
    \item The chromo-magnetic field $\vec{B}$ induced by the static sources is statistically consistent with zero within our present accuracy.  The chromo-electric field $\vec{E}$ is a sum of a nonperturbative longitudinal field $\vec{E}^{\rm NP}$, \footnote{In what follows, for brevity we omit the prefix ``chromo-'' when referring to field components.} aligned with the interquark axis,  and a perturbative irrotational field, $\vec{E}_{\rm C}$.  
    The field $\vec{E}^{\rm NP}$
   is responsible for confinement and manifests itself as a smooth, spatially extended flux tube connecting the two static sources.

    \item The curl of the electric field defines a magnetic current $\vec{J}_{\rm mag}$ that encircles the flux-tube axis and exhibits continuum scaling.
The current produces a Lorentz force density  
    \[
    \vec{f} \;=\; \vec{J}_{\rm mag} \times \vec{E}^{\rm NP},
    \]

    where both $\vec{J}_{\rm mag}$ and $\vec{E}^{\rm NP}$ are determined directly from lattice simulations. This relation provides nontrivial support for the Maxwell picture of confinement~\cite{Baker:2019gsi}, according to which confinement arises through the dual-superconductor mechanism, with electric flux tubes stabilized by circulating magnetic currents. In addition, theoretical analyses such as those of Ref.~\cite{Cea:2023} indicate that the longitudinal electric field is predominantly governed by its Abelian components, thereby further substantiating the dual-superconductor scenario.
\end{itemize}

\begin{figure*}[htb]
\begin{center}
\includegraphics[width=0.47\linewidth,clip]{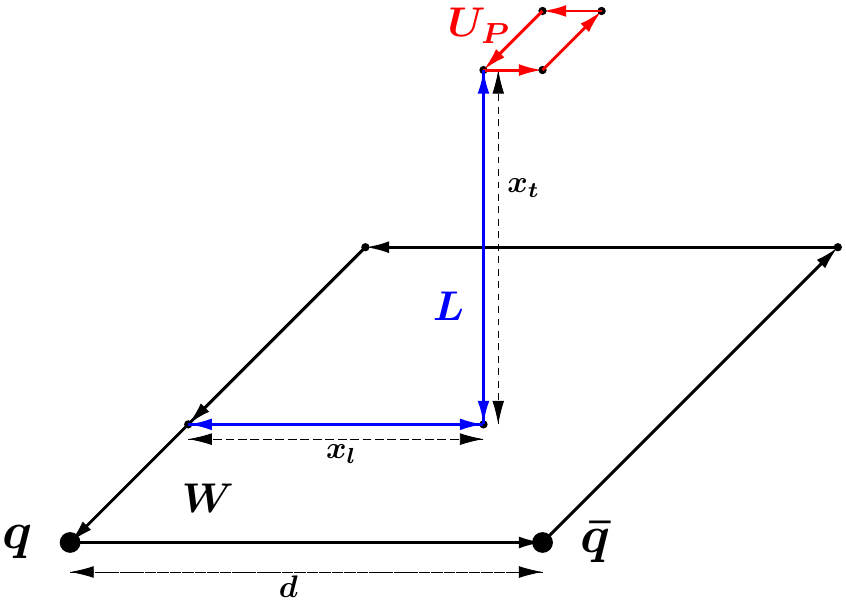}
\includegraphics[width=0.47\linewidth,clip]{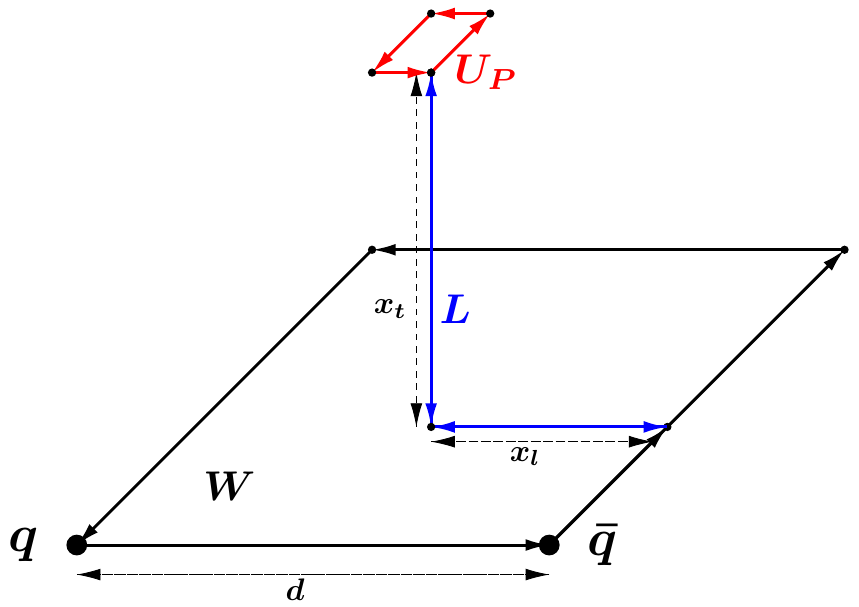}
\end{center}
\caption{The flux-tube operator with the Schwinger line attached to the quark time line (left) or to the antiquark time line (right).
}            
\label{fluxtubeoperator}
\end{figure*}

In the present work we investigate full QCD with \(2{+}1\) HISQ flavors at finite temperature. The light and strange quark masses are tuned along the line of constant physics corresponding to a physical pion mass of \(m_\pi = 140\) MeV (see Ref.~\cite{Bazavov:2017dus}). The primary goal of this study is to analyze the thermal behavior of flux-tube structures generated by a static quark--antiquark pair. \\
Strong interacting matter as described by QCD exhibits two important 
 nonperturbative features, namely spontaneous breaking of the chiral symmetry and color confinement.  It is  widely believed that, 
 at a sufficiently high temperature, QCD undergoes a phase transition where quarks and gluons are deconfined in a quark-gluon plasma and the
 chiral symmetry is restored. Usually, the order parameters employed for chiral and deconfinement phase transitions are the light-quark chiral
condensate and the Polyakov loop. Indeed, the Polyakov loop, which is related to the free energy of a static color charge as probe of the system,
should  offer a clear physical picture of the deconfinement transition. It turned out that when the bare chiral condensate starts to decrease
towards chiral symmetry restoration, the expectation value of the bare Polyakov loop starts to increase, signaling the onset of deconfinement. Moreover,
 the chiral critical temperature $T_{\rm c}$ and deconfinement critical temperatures $T_{\rm dec}$, as estimated from
the inflection points of the chiral condensate and Polyakov loops, are almost coincident. 
However, both the bare chiral condensate and Polyakov loop are affected by ultraviolet divergencies 
 that need to be renormalized to achieve a sensible continuum limit.  So that, usually, the order parameters employed 
for chiral and deconfinement phase transitions are the renormalized light chiral condensate and Polyakov loop.
It is, now,  well established that for physical quark masses the transition between the hadronic and the quark-gluon phases
is an analytic crossover~\cite{Aoki:2006we,Bhattacharya:2014ara}. In general, in an analytic transition, due to the absence of singular behavior, 
a unique critical temperature cannot be defined, nevertheless  a pseudocritical temperature can be determined using the inflection point
or peak position of certain thermodynamic observables.  As for the chiral pseudocritical temperature, to date the most precise determination
has been reported in Ref.~\cite{HotQCD:2018pds}, where the  pseudocritical temperature of the QCD chiral crossover was obtained from lattice QCD calculations carried out with two degenerate up and down dynamical quarks and a dynamical strange quark, with quark masses corresponding to physical values of pion and kaon masses in the continuum limit:
\begin{equation}
\label{1.1}
   T_{\rm c} \;  = \; 156.5 \; \pm \; 1.5 \; \; \text{MeV} \;  \; .
\end{equation}
On the other hand, the fact that  the QCD transition is a nonsingular crossover implies that  different observables lead 
to different numerical values for the pseudocritical temperature,  even in the  continuum  and thermodynamic limits.
In fact, Ref.~\cite{Aoki:2006br}  reported that the peak of the renormalized Polyakov loop susceptibility resulted in 
 a pseudocritical temperature  25(4)  MeV larger  with respect to  the peak of the renormalized chiral susceptibility. 
Another consequence of the crossover is the nonvanishing width of the peaks even in the thermodynamic limit.
In any case, there is an almost unanimous consensus on the fact that the crossover from hadronic to quark matter 
should be  both a  deconfinement as well as a chiral-symmetry-restoring transition with $T_{\rm dec} \; \sim  \; T_{\rm c}$, or more 
precisely,  $|T_{\rm dec} \; -  \; T_{\rm c} |\; \ll \; T_{\rm c}$, where we stress that  here $T_{\rm dec}$ indicates the pseudocritical temperature extracted from
the renormalized Polyakov loop (a good account can be found in the recent reviews Refs.~\cite{Gross:2022hyw,Aarts:2023vsf} 
and references therein).\\

In our previous papers (see Ref.~\cite{Baker:2024peg} and references therein) we have seen how  our connected correlation function
allows us to study the detailed structure of the color fields in the region around quark and antiquark static sources at zero temperature.
 The resulting Maxwell picture of the flux tube stabilized by circulating magnetic current $\vec{J}_{\rm mag}$ leads to a well-defined mechanism for the squeezing of the field $\vec{E}$ into a narrow flux tube.
The magnetic current is almost uniform along the flux tube and it decreases rapidly in the transverse directions, thus assuring
the presence of an almost uniform and narrow longitudinal field along the flux tube. 
 
In this paper we are interested in the behavior of the flux tubes for full QCD near the deconfinement  chiral transition and we extend this analysis to finite temperatures, employing both quark and antiquark operators (see the discussion at the end of Section 2).

The rest of this paper is organized as follows. In Section~2 we recall the theoretical background underlying our approach. Section~3 describes the setup of our numerical simulations, including the lattice action, scale setting, and smearing procedure. In Section~4 we present our numerical results, while Section~5 is devoted to their discussion. Finally, in Section~6 we summarize our conclusions.
%

%
%
\section{The field strength tensor}

In close analogy with our previous analyses~\cite{Baker:2018mhw,Baker:2019gsi,Baker:2022cwb,Baker:2024peg}, we extract the spatial distribution of the color fields generated by a static quark--antiquark pair from lattice measurements of the connected correlation function \(\rho^{\text{conn}}_{W,\mu\nu}\)~\cite{DiGiacomo:1989yp}. This correlator involves a plaquette \(U_P = U_{\mu\nu}(x)\) in the \(\mu\nu\) plane and a rectangular Wilson loop \(W\) (see Fig.~\ref{fluxtubeoperator}, left),
\begin{equation}
    \rho^{\text{conn}}_{W,\mu\nu} = 
    \frac{\langle \mathrm{tr}(W L U_P L^\dagger)\rangle}{\langle \mathrm{tr}(W)\rangle}
    - \frac{1}{N}\,\frac{\langle \mathrm{tr}(U_P)\,\mathrm{tr}(W)\rangle}{\langle \mathrm{tr}(W)\rangle}\;,
    \label{connected1}
\end{equation}
where \(N=3\) is the number of QCD colors. The correlator \(\rho^{\text{conn}}_{W,\mu\nu}\) provides a gauge-invariant lattice definition of the field-strength tensor induced by the static sources, 
\(\langle F_{\mu\nu}\rangle_{q\bar q} \equiv F_{\mu\nu}\), 
which carries one unit of octet charge while preserving the space--time symmetry properties of the Maxwell tensor in electrodynamics:
\begin{equation}
    \rho^{\text{conn}}_{W,\mu\nu} \;\equiv\; a^2 g \langle F_{\mu\nu}\rangle_{q\bar q} 
    \;\equiv\; a^2 g\,F_{\mu\nu}\;.
    \label{connected2}
\end{equation}
Depending on the orientation of the plaquette, different field components are accessed. For instance, if \(U_P\) lies in the \(\hat{4}\hat{1}\) plane, the measured component \(F_{41}\) corresponds to \(E_x\), the longitudinal electric field along the quark--antiquark axis, evaluated at the plaquette center. If the plaquette is in the \(\hat{4}\hat{2}\) plane, one measures \(F_{42} = E_y\), a transverse component of the electric field with respect to the \(q\bar q\) axis. Similarly, for a plaquette in the \(\hat{2}\hat{3}\) plane one obtains \(F_{23} = B_x\), the longitudinal magnetic component, and so on for the other orientations.
The Maxwell-like field tensor \(F_{\mu\nu}\), defined in Eq.~\eqref{connected2}, induces a magnetic current density \(J_\alpha^{\rm mag}\) circulating around the flux-tube axis:  
\begin{equation}
     J_\alpha^{\rm mag} \equiv \tfrac{1}{2}\,\epsilon_{\alpha \beta \mu \lambda}\, 
     \frac{\partial F_{\mu \lambda}}{\partial x^\beta}, 
     \qquad (\epsilon_{4123} = 1)\; .
     \label{Jmag}
\end{equation}
The presence of magnetic currents in \(\mathrm{SU(3)}\) lattice gauge theory was already anticipated in Ref.~\cite{Skala:1996ar}.  

The corresponding force density acting on these currents is determined by the Maxwell stress tensor,  
\begin{equation}
 T_{\alpha \beta} = F_{\alpha \lambda} F_{\beta \lambda} 
 - \tfrac{1}{4}\,\delta_{\alpha \beta}\,F_{\mu \lambda} F_{\mu \lambda}\ , 
 \label{stress}   
\end{equation}
through
\begin{equation}
    f_\beta = \frac{\partial}{\partial x^\alpha} T_{\alpha \beta}
    = - F_{\mu \lambda}\,\tfrac{1}{2}\,\epsilon_{\alpha \beta \mu \lambda}\, J^\text{mag}_\alpha \; .
    \label{Fs}
\end{equation}

If the magnetic components of the field tensor, \(\tfrac{1}{2}\epsilon_{ijk} F_{jk}\), vanish, Eq.~\eqref{Jmag} reduces in its spatial part to  
\begin{equation}
  \vec{J}_\text{mag} = \vec{\nabla} \times \vec{E}\ ,
  \label{rotel}
\end{equation}
while Eq.~\eqref{Fs} becomes
\(\vec{f} = \vec{J}_\text{mag} \times \vec{E}\). 

The general solution of  Eq.~\eqref{rotel} is given by
\begin{equation}
\label{1.3}
 \vec{E}(\vec{x}) \; = \; \vec{E}^{\rm NP}(\vec{x})     \;  + \;  \vec{E}_{\rm C}(\vec{x}) \; \; ,
\end{equation}
where $\vec{E}_{\rm C}(\vec{x})$ is a perturbative irrotational field and 
 $\vec{E}^{\rm NP}(\vec{x})$ is the nonperturbative longitudinal field satisfying 
\begin{equation}
\label{1.4}
\vec{\nabla} \; \times \vec{E}^{\rm NP}(\vec{x}) \; = \; \vec{J}_{\rm mag}(\vec{x})     \;  \; .
\end{equation}

 We have developed a model-independent analysis  to evaluate 
$\vec{E}_{\rm C}(\vec{x})$ by numerically solving $\vec{\nabla} \; \times \vec{E}_{\rm C}(\vec{x}) \, = \, 0$ and to determine $\vec{E}^{\rm NP}$
 by numerical solution of \cref{1.4}. Moreover, by numerically evaluating the curl of the full electric
field, we were able  to track the transverse distribution of the magnetic current in a model-independent way.

Replacing the electric field \(\vec{E}\) by its nonperturbative longitudinal component \(\vec{E}^{\rm NP}\), as determined from our lattice measurements of the connected correlator, we finally obtain  $\vec{f}$, the confining force density directed toward the flux-tube axis;
\begin{equation}   
  \vec{f} = \vec{J}_{\rm mag} \times \vec{E}^{\rm NP}\ .
  \label{vecfdensity}
\end{equation}
  
Thus, we see that model-independent evidence of confinement can be  obtained from the magnetic current $\vec{J}_{\rm mag}$ and the nonperturbative
electric field $\vec{E}^{\rm NP}$ generated by a static quark-antiquark pair.
This analysis was implemented in Ref.~\cite{Baker:2024peg} for full
QCD at almost zero temperatures, {\it i.e.} in the confined phase.

This ``Maxwell'' picture of confinement is further supported by recent studies of the \(\mathrm{SU(3)}\) gauge theory~\cite{Cea:2023}, showing that the electric field \(gE^a\) inside the flux tube is predominantly carried by its Abelian components \(gE^3\) and \(gE^8\).  \\
At zero temperature, the spatial distributions of the color fields induced by a static quark-antiquark pair can be obtained from lattice measurements of the connected correlation function \cref{connected1}. On the other hand,
at nonzero temperatures, the operator $\rho^\text{conn}_{W, \mu \nu}$ should be replaced by~\cite{Skala:1996ar}
\begin{equation}
    \rho^\text{conn}_{P, \mu \nu} = \frac {\langle\tr (PLU_PL^*)\tr(P^\dagger)\rangle}{\langle\tr(P)\tr(P^\dagger)\rangle} - \frac{1}{N} \frac {\langle\tr (U_P) \tr (P)\tr(P^\dagger)\rangle}{\langle\tr(P)\tr(P^\dagger)\rangle}\;,
    \label{connected1P}
\end{equation}
where the $P$ and $P^\dagger$ denote two parallel Polyakov lines with opposite orientations, separated by the distance $d$. The lattice operator $\rho^\text{conn}_{P, \mu \nu}$ was adopted in some previous studies of SU(3) pure gauge theory at zero temperature~\cite{Cea:2013oba,Cea:2014uja,Cea:2014hma} and at nonzero temperature~\cite{Cea:2015wjd}.
It turned out, however, that even when $\rho^\text{conn}_{P, \mu \nu}$ is measured on smeared ensembles, there was a strong asymmetry between the quark and antiquark
static color sources. For this reason in Ref.~\cite{Baker:2023dnn}, instead of 
$\rho^\text{conn}_{P, \mu \nu}$, a connected correlation function  was  adopted involving maximal Wilson loops, {\it i.e.} loops with the largest possible extension 
in the temporal direction. So that, in the present paper, the connected lattice  operator
 \cref{connected1} always refers to the maximal Wilson loop. It is worth stressing that
the operators $\rho^\text{conn}_{W_{\rm max}, \mu \nu}$ and 
$\rho^\text{conn}_{P, \mu \nu}$ differ in a sense which can be understood considering their counterparts when the Schwinger line and the plaquette are removed. As shown in Ref.~\cite{Jahn:2004qr}, a standalone maximal Wilson loop with a given extension $d$ in the spatial direction is equivalent to the singlet correlator of two Polyakov loops, 
$\langle {\rm tr}(P P^\dagger)\rangle$, at distance $d$, calculated in the axial gauge,
 on a lattice with periodic boundary conditions, and gives
access to the static potential for a quark-antiquark pair in the singlet state; instead, a gauge-invariant  Polyakov loop correlator  gives access to a combination of the quark-antiquark static potential in the singlet and octet states. \\
Even using the  connected correlation function with maximal Wilson loop operator, we noticed that  small quark-antiquark asymmetries were still persistent in the lattice data.
For this reason we decided to measure also the connected correlation function where
the Schwinger line is connected to the antiquark time line (see Fig.~\ref{fluxtubeoperator}, right).
As we will show later on, in this way our lattice measurements display an almost
perfect quark-antiquark symmetry for the longitudinal electric field $E_x$ and anti-symmetry
for the transverse electric fields $E_y$ and $E_z$. In addition, we checked that 
on the midplane of the static color sources both quark and antiquark connected 
correlation functions returned  electric fields consistent within the statistical uncertainties. This last point is a nontrivial consistency check, demonstrating that
the field-strength tensor does not depend on the  Schwinger-line path. \\
\section{Lattice setup}

\begin{figure*}[htb]
\begin{center}
\includegraphics[width=0.9\linewidth,clip]{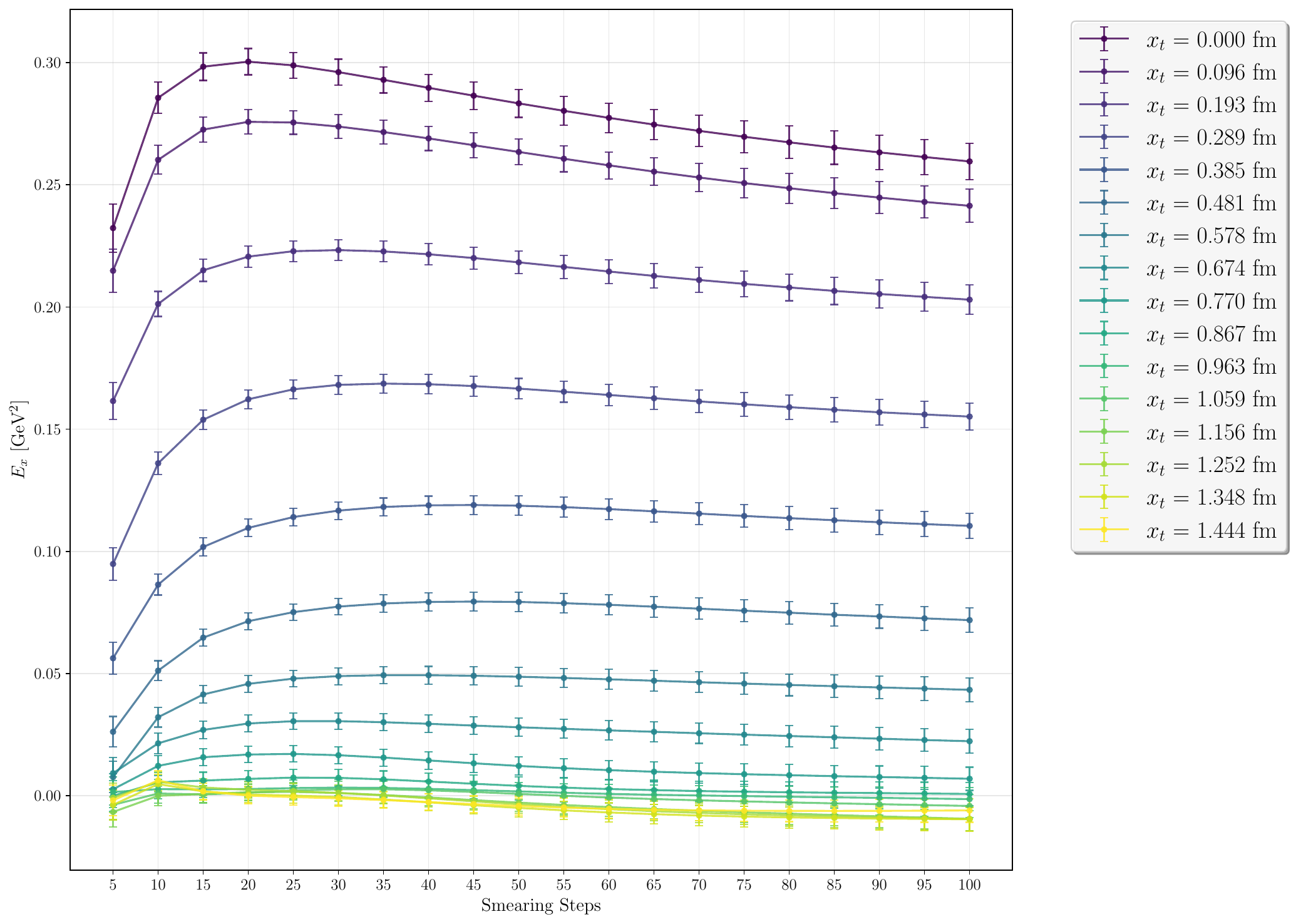}
\end{center}
\caption{The $E_x$ field component for $x_l=4a \simeq 0.385$ fm, at $\beta=6.880$, $48^4$ lattice, $d \simeq 0.963$ fm, {\it versus} HYP3d smearing steps.}
\label{smearingquarkside}
\end{figure*}

\begin{figure*}[htb]
\begin{center}
\includegraphics[trim=-8.685 -2.111 -1.903 3.625,width=0.47\textwidth,clip]{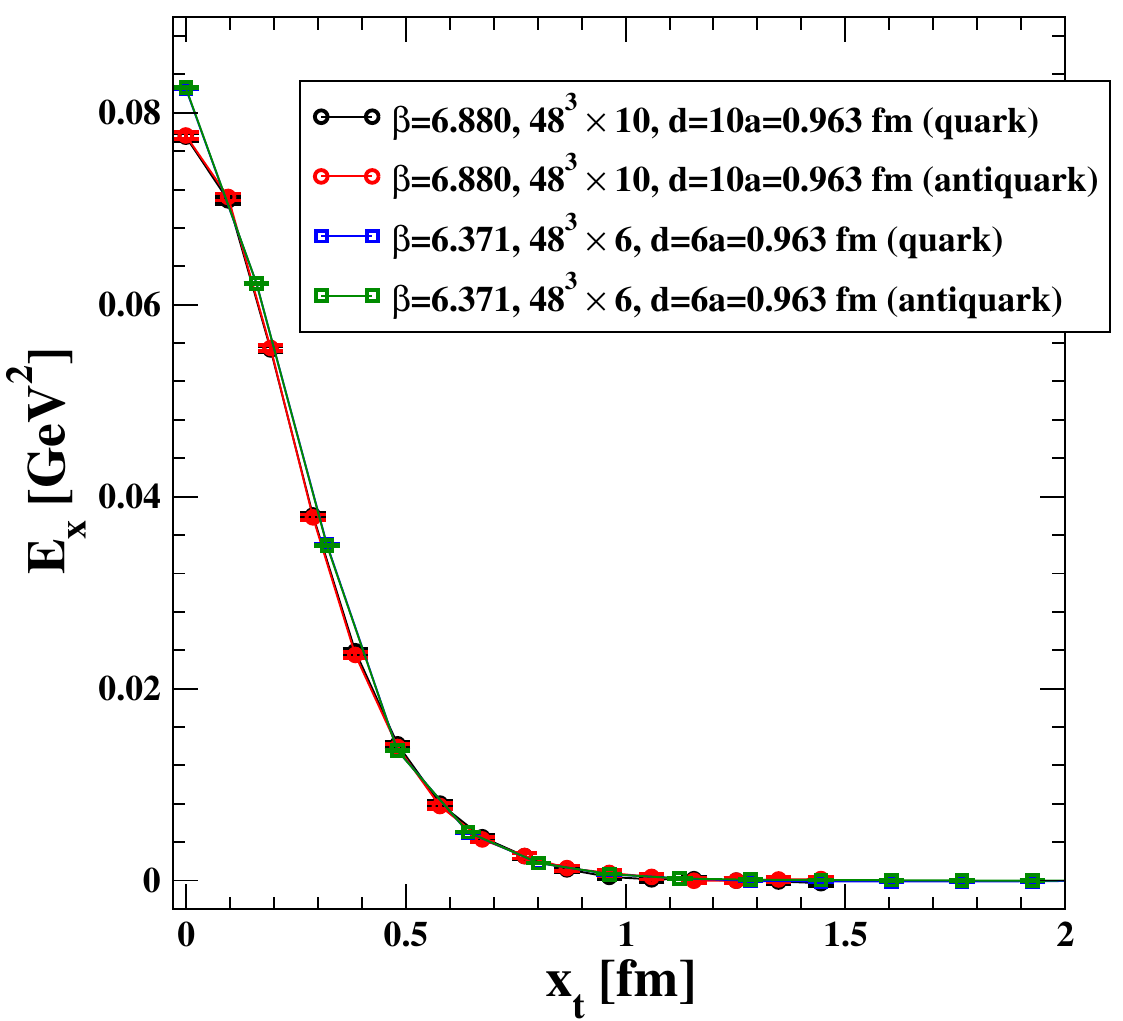}
\includegraphics[width=0.47\textwidth,clip]{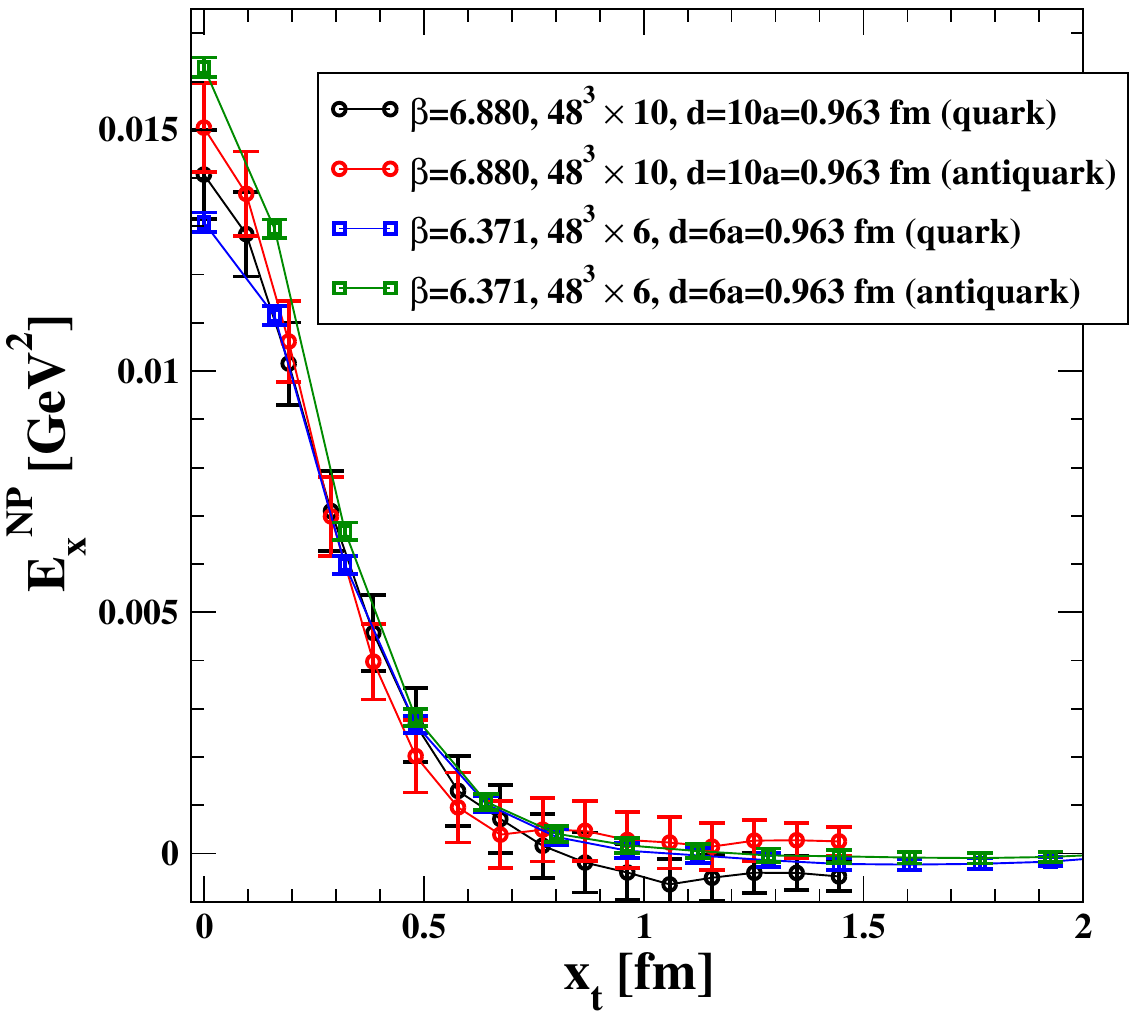}
\end{center}
\caption{Scaling check at $T \simeq 205$ MeV and $d \simeq 0.963$ fm for the full longitudinal electric field $E_x$ (left) and its nonperturbative component $E_x^{\rm NP}$ (right).
}            
\label{scalingT205MeV}
\end{figure*}

\begin{figure*}[htbp]
\begin{center}
\includegraphics[width=0.47\linewidth,clip]{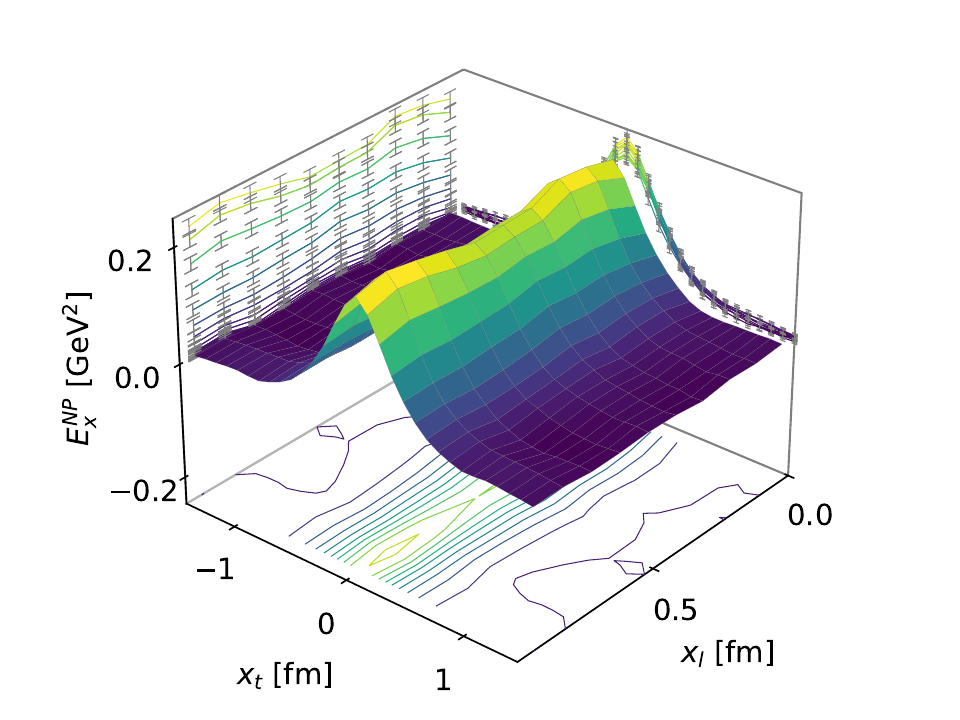}
\includegraphics[width=0.47\linewidth,clip]{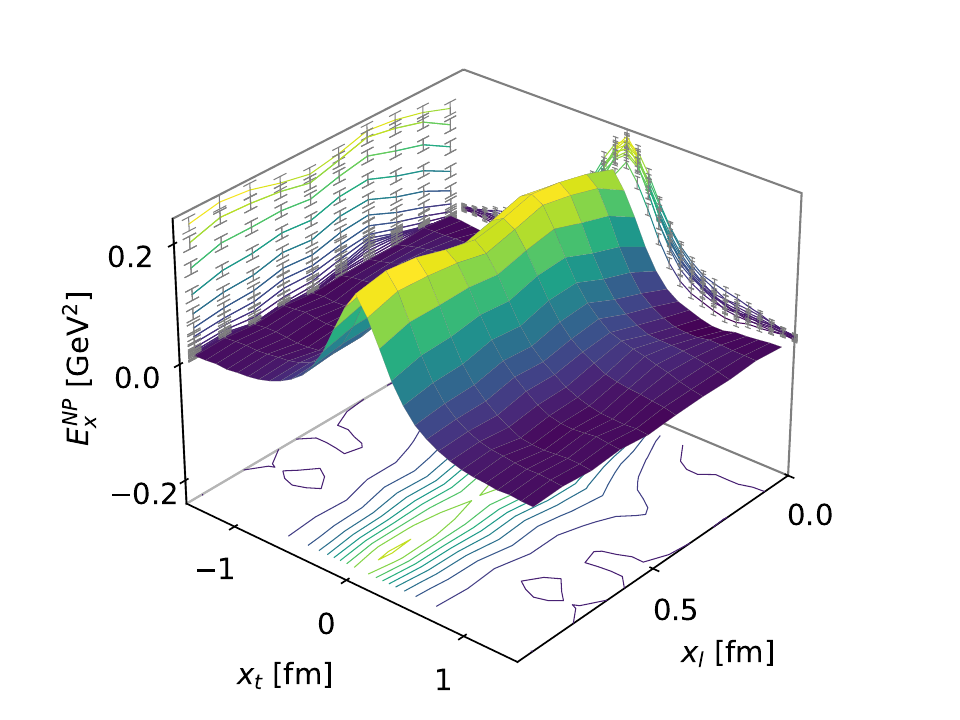}
\\
\includegraphics[width=0.47\linewidth,clip]{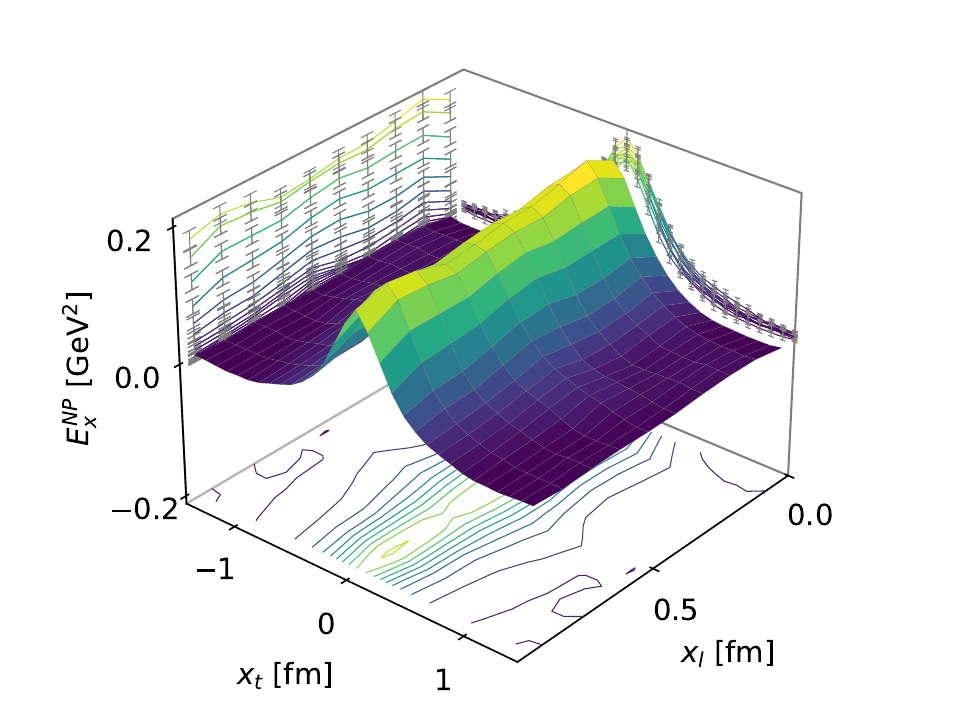}
\includegraphics[width=0.47\linewidth,clip]{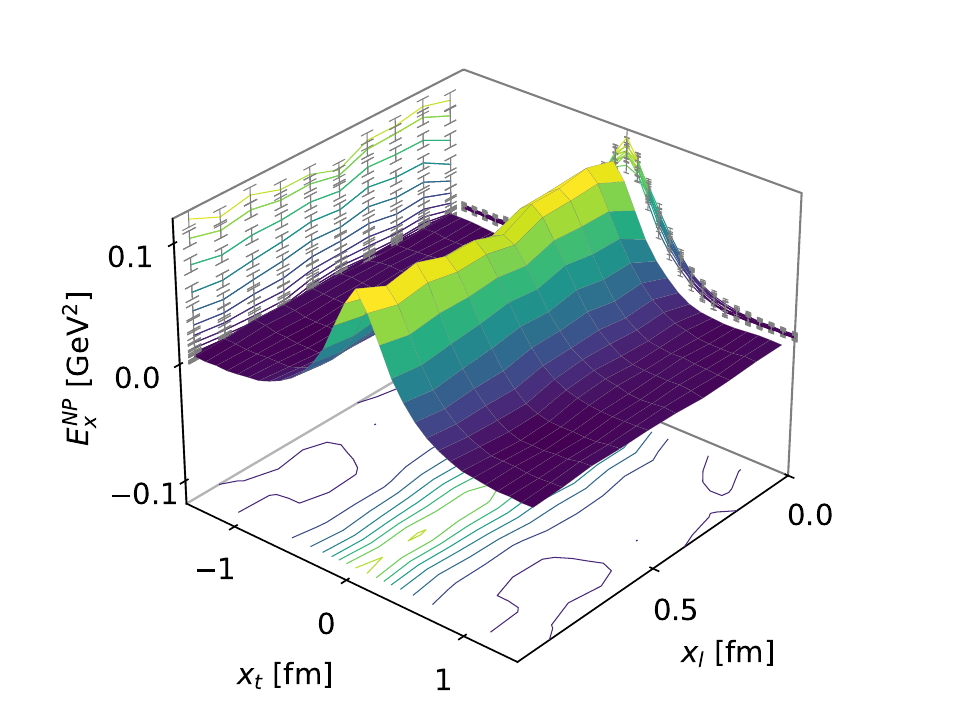}
\\
\includegraphics[width=0.47\linewidth,clip]{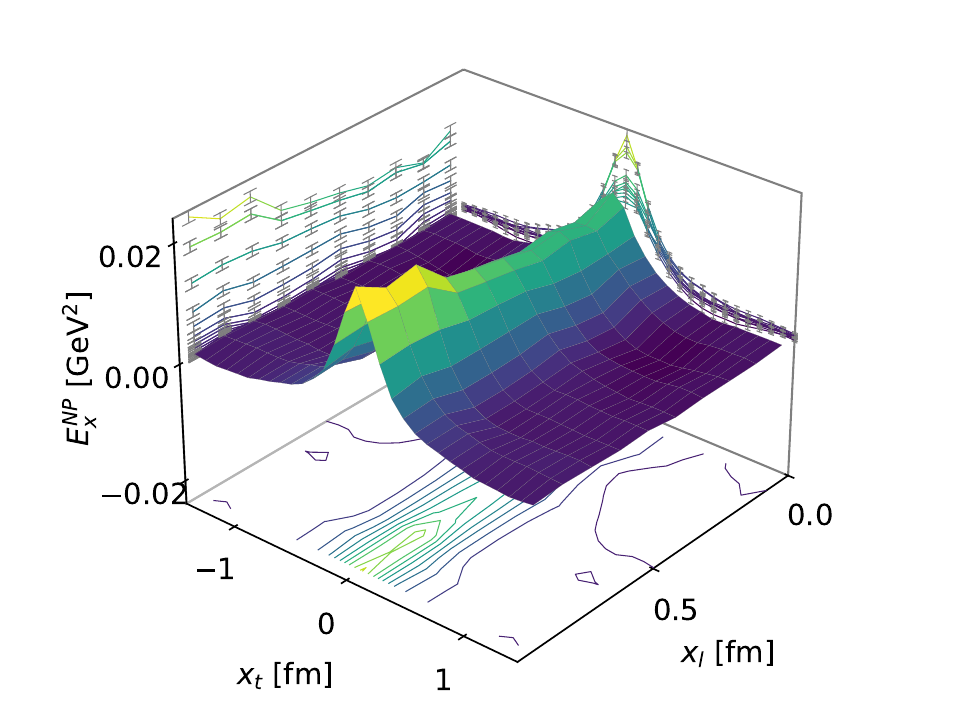}
\includegraphics[width=0.47\linewidth,clip]{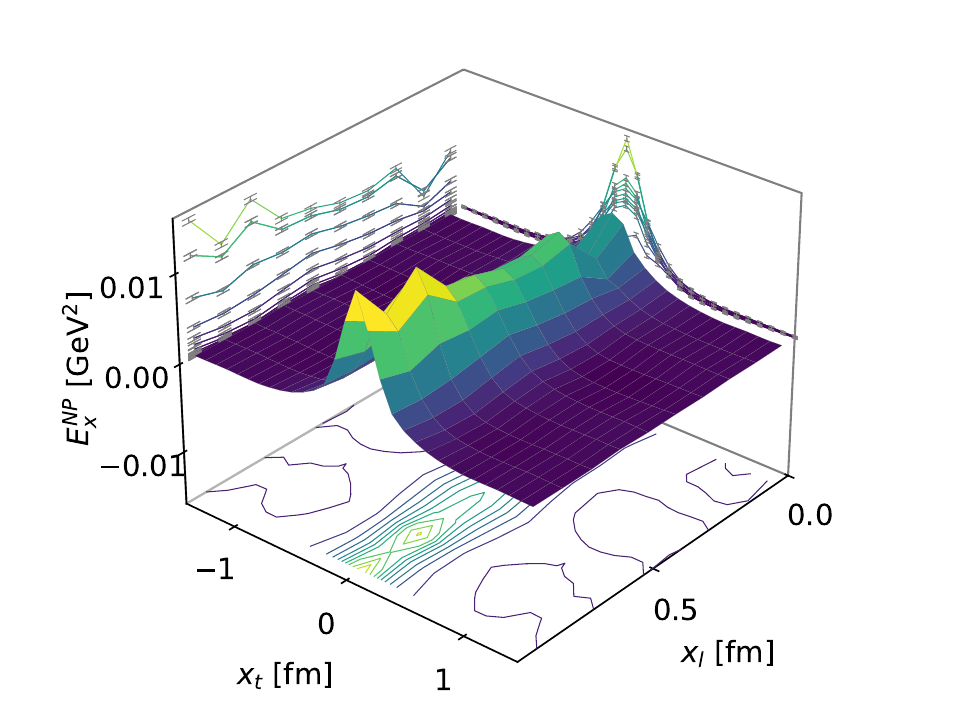}
\\
\includegraphics[width=0.47\linewidth,clip]{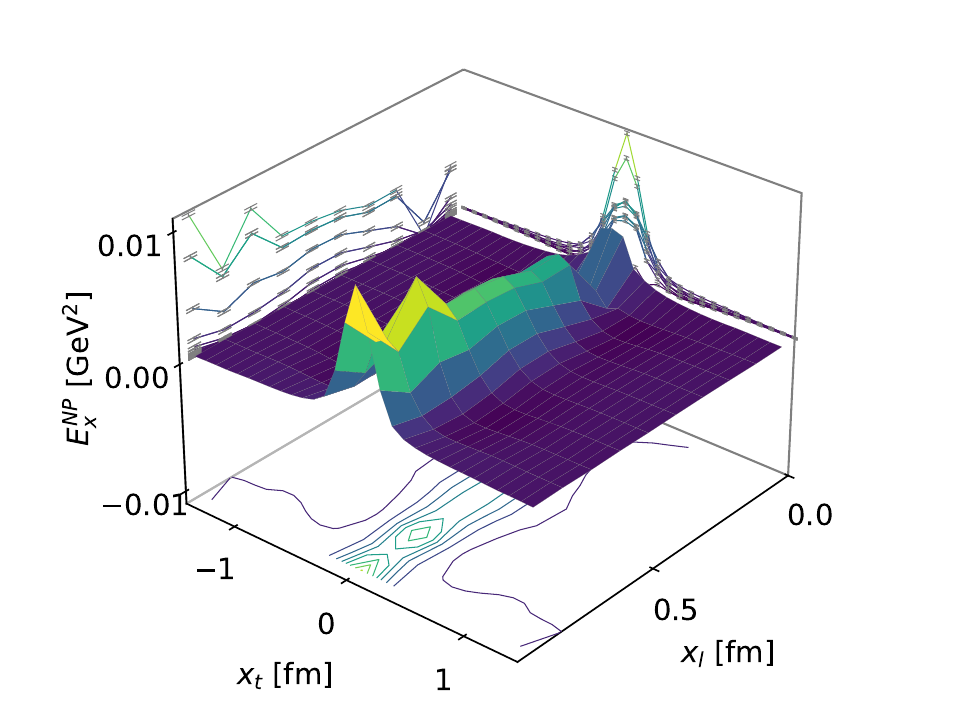}
\includegraphics[width=0.47\linewidth,clip]{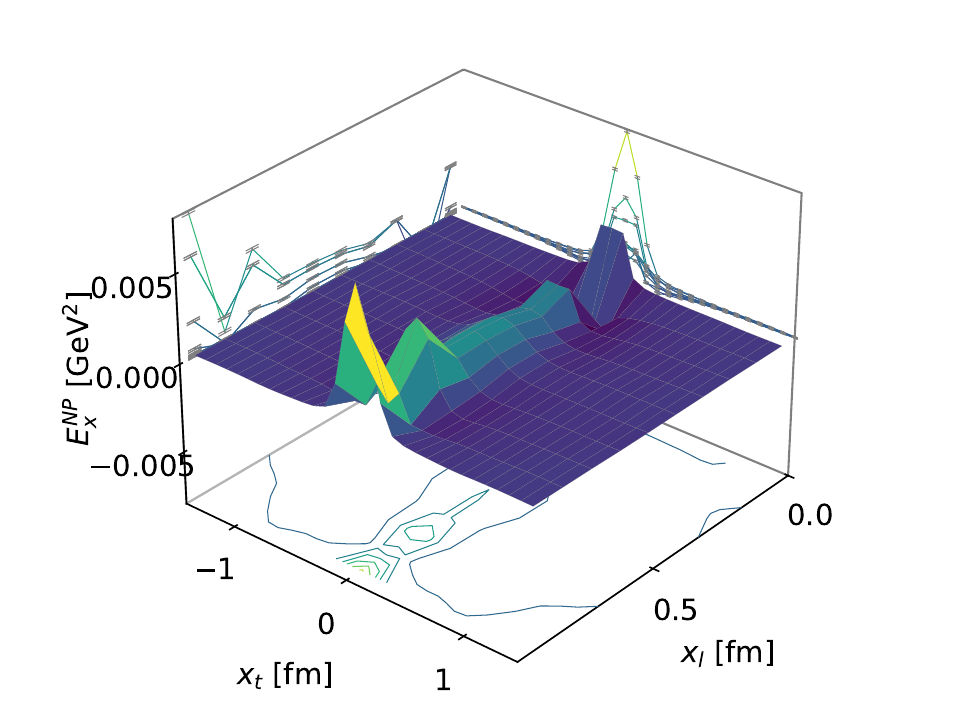}
\end{center}
\caption{The full profile of $E_x^{\rm NP}$ field at $\beta=6.880$, $d=10a \simeq 0.963$ fm, in correspondence of increasing temperatures: $T \simeq 43,128,146,171,205,256,342,512$ MeV (from left to right, from top to bottom).
}            
\label{ExNP3d_merged}
\end{figure*}

\begin{figure*}[htbp]
\begin{center}
\includegraphics[width=0.95\linewidth,clip]{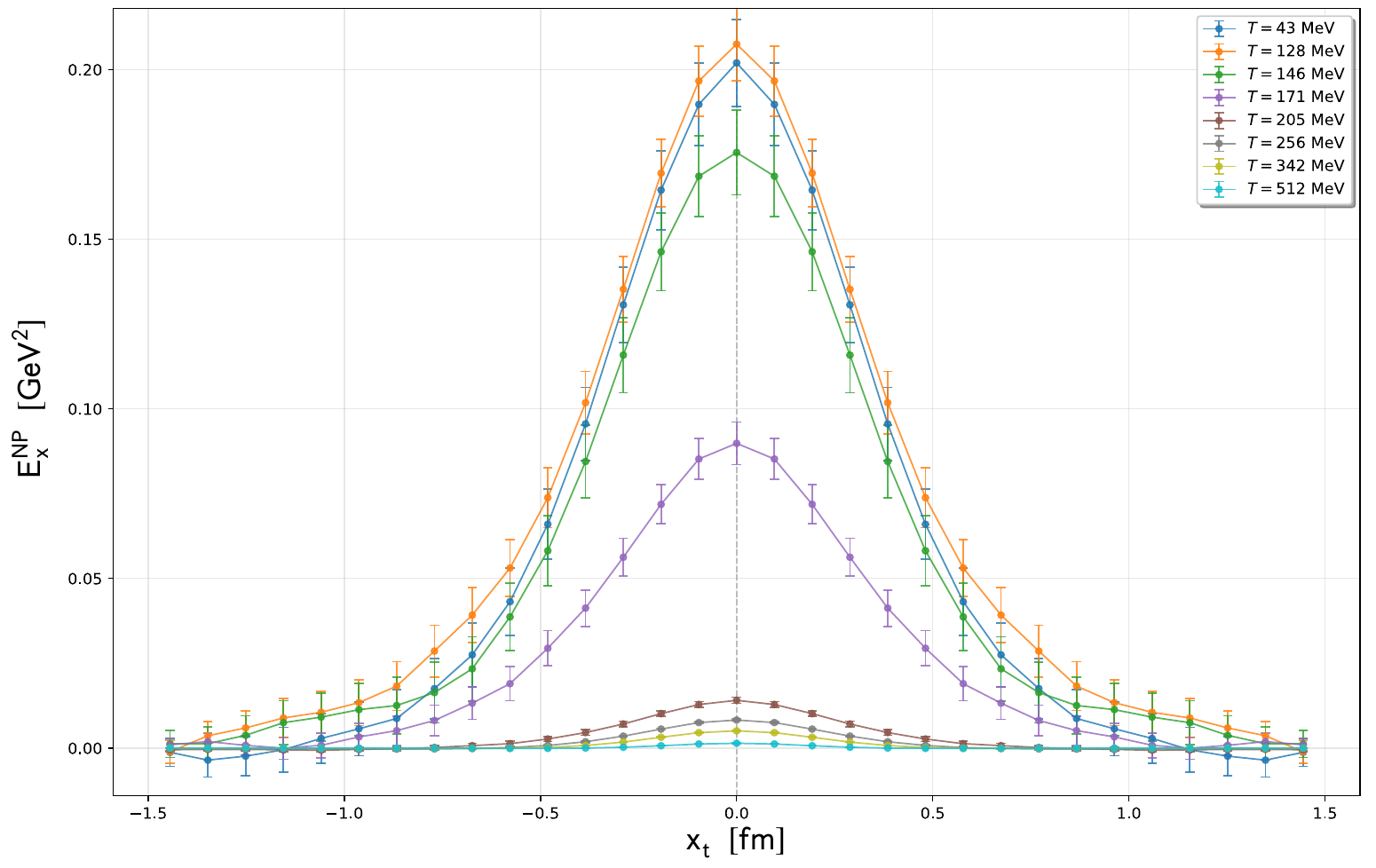}
\end{center}
\caption{$E_x^{\rm NP}$ field on the transverse midplane for increasing temperatures: $T \simeq 43,128,146,171,205,256,342,512$ MeV .
}            
\label{ExNP_midpoint_vs_T}
\end{figure*}

\begin{table*}[th]
\begin{center} 
  \caption{Summary of the measurements.}
  \label{measurements}
\setlength{\tabcolsep}{20pt}
\begin{tabular}{cccrrcc}
\toprule
lattice             & $\beta=10/g^2$ & $a(\beta)$ [fm]   &  $T$ [\textrm{MeV}]  &  $d$ [lattice units]  &   $d$ [fm]      & measurements          \\ \midrule

$ 48^4           $	&	6.880	     & 0.0963  &	43	\hspace{0.2cm}    &   10    \hspace{0.6cm}            &   0.963       &   1635               \\
$ 48^3 \times 16 $	&	6.880	     & 0.0963  &	128  \hspace{0.2cm}          &   10  \hspace{0.6cm}              &   0.963       &   6253               \\
$ 48^3 \times 14 $	&	6.880	     & 0.0963  &	146  \hspace{0.2cm}            &   10   \hspace{0.6cm}             &   0.963       &   3605               \\
$ 48^3 \times 12 $	&	6.880	     & 0.0963  & 	171   \hspace{0.2cm}           &    6   \hspace{0.6cm}             &   0.578       &   3707               \\
$ 48^3 \times 12 $  &   6.880        & 0.0963  &    171  \hspace{0.2cm}            &    8   \hspace{0.6cm}             &   0.770       &   3214               \\
$ 48^3 \times 12 $  &   6.880        & 0.0963  &    171    \hspace{0.2cm}          &   10   \hspace{0.6cm}             &   0.963       &   3707               \\
$ 48^3 \times 12 $  &   6.880        & 0.0963  &    171   \hspace{0.2cm}           &   12   \hspace{0.6cm}             &   1.156       &   3709               \\
$ 48^3 \times 12 $  &   6.880        & 0.0963  &    171  \hspace{0.2cm}            &   14   \hspace{0.6cm}             &   1.348       &   3712               \\
$ 48^3 \times 10 $	&	6.880	     & 0.0963  &	205   \hspace{0.2cm}           &   10   \hspace{0.6cm}             &   0.963       &   5424               \\
$ 48^3 \times 8  $	&	6.880	     & 0.0963  &	256   \hspace{0.2cm}           &   10   \hspace{0.6cm}             &   0.963       &   5028               \\
$ 48^3 \times 6  $	&	6.880	     & 0.0963  &	342   \hspace{0.2cm}           &   10   \hspace{0.6cm}             &   0.963       &   4685               \\
$ 48^3 \times 4  $	&	6.880	     & 0.0963  &	512    \hspace{0.2cm}          &   10   \hspace{0.6cm}             &   0.963       &   5395               \\
$ 48^3 \times 6  $	&	6.371        & 0.1605  &    205    \hspace{0.2cm}          &    6   \hspace{0.6cm}             &   0.963       &   4501               \\

\bottomrule 
\end{tabular}
\end{center}
\end{table*}

We perform simulations of lattice QCD with 2+1 flavors of HISQ (Highly Improved Staggered Quarks) 
quarks. We have made use of the HISQ/tree action~\cite{Follana:2006rc,Bazavov:2009bb,Bazavov:2010ru}.
Couplings are adjusted so as to move on a line of constant physics (LCP), as determined in Ref.~\cite{Bazavov:2011nk}, with the strange quark mass $m_s$ fixed at its physical value and a light-to-strange mass ratio $m_l/m_s=1/27$, corresponding to a pion mass of 140 MeV in the continuum limit.
We have simulated the theory for several values of 
the gauge coupling, adopting lattices of size  $48^3\times L_t$,  $4 \le L_t \le 48$ (see Table~\ref{measurements}). 
The thermalized lattice configurations, stored for further measurements, are each separated by 25 trajectories of rational hybrid Monte Carlo (RHMC) with length one.
The electromagnetic field tensor has been measured for a static quark and antiquark placed at a given distance $d$.\
We fix the lattice spacing through the  observable $f_K$ as defined in the Appendix~B of  Ref.~\cite{Bazavov:2011nk}.

The lattice spacing in physical units is given by~\cite{Bazavov:2011nk,MILC:2010hzw}
\begin{equation}
\label{abeta}
a(\beta) \;=\; \frac{r_1}{r_1 f_K} \frac{c_0^K f(\beta)+c_2^K (10/\beta) f^3(\beta)}{
1+d_2^K (10/\beta) f^2(\beta)} \;,
\end{equation}
where
\begin{equation}
\label{r1fkappa}
\begin{split}
r_1 \;  &= \; 0.3106 \; \mathrm{fm} \;,
\\
 r_1 f_K \;  &= \; 
\frac{0.3106 \; \mathrm{fm} \; \cdot \; 156.1/\sqrt{2} \; \mathrm{MeV}}{
197.3 \; \mathrm{MeV \; fm}} \; \;
\end{split}
\end{equation}
and
 \begin{equation}
\label{coefficients}
c_0^K = 7.66 \; \;  , \; \;  c_2^K \; = \; 32911 \; \; , \; \;  d_2^K  \;  =  \;  2388 \; \; .
\end{equation}
In Eq.~(\ref{abeta})  $f(\beta)$ is the two-loop beta  function:
\begin{equation}
\label{3.6}
f(\beta)=[b_0 (10/\beta)]^{-b_1/(2 b_0^2)} \exp[-\beta/(20 b_0)] \; ,
\end{equation}
$b_0$ and $b_1$ being its universal coefficients. 

Our setup employs one step of four-dimensional hypercubic smearing (HYPt) on the temporal links, with smearing parameters \((\alpha_1,\alpha_2,\alpha_3) = (1.0, 1.0, 0.5)\)~\cite{Hasenfratz:2001hp}. In addition, we apply \(N_{\rm HYP3d}\) steps of hypercubic smearing restricted to the three spatial directions (HYP3d), using parameters \((\alpha_1^{\text{HYP3d}},\alpha_3^{\text{HYP3d}}) = (0.75, 0.3)\).  
Figure~\ref{smearingquarkside} illustrates the typical behavior of the \(E_x(x_l,x_t)\) component of the tensor field under HYP3d smearing. The numerical results shown correspond to a \(48^4\) lattice at \(\beta=6.880\), with longitudinal separation \(x_l = 4a \simeq 0.385\) fm. As seen in the figure, 50 HYP3d smearing steps are sufficient to stabilize the results for all transverse distances \(x_t\) in the range \((0.000,\,1.444)\) fm.  
Accordingly, in the remainder of this paper we report measurements performed after 50 HYP3d smearing steps. We notice that the smearing strategy adopted in this paper is different with respect to our previous papers, where we tuned the smearing step, for each value of the transverse coordinate $x_t$, at the ``optimal" value where the observable under consideration reaches the maximum. We have checked that the two procedures, for the lattice setups considered in the present work, do not lead to significant differences.

In Table~\ref{measurements} we display the list of measurements we have done. We considered distances between the sources in a wide interval, ranging from 0.578~fm to 1.348~fm, corresponding to $\beta$-values in the range 6.371 to 6.880. In this and following tables we round the value of temperature to the nearest integer (in MeV), and the values of distances to three decimal digits (in fm).

\section{Numerical results}

\subsection{Scaling check}
\label{scaling}
We have verified that our lattice setup is sufficiently close to the continuum limit by checking that different choices of lattice parameters, corresponding to the same physical distance \(d\) between the sources and the same temperature \(T\), yield consistent values of the relevant observables when expressed in physical units.  

In Fig.~\ref{scalingT205MeV} we show the case with source separation \(d \simeq 0.963\) fm at temperature \(T \simeq 205\) MeV. To realize these physical values of distance and temperature, we employed two values of the gauge coupling, \(\beta=6.371\) and \(\beta=6.880\), spanning a fairly wide interval. We compared the physical values of the full longitudinal electric field \(E_x\) and its nonperturbative component \(E_x^{\mathrm{NP}}\).  

As an additional consistency check, we measured \(E_x\) and \(E_x^{\mathrm{NP}}\) using the operator of Eq.~(\ref{connected1}) with the Schwinger line attached either to the quark temporal line or to the antiquark temporal line. The numerical results, displayed in Fig.~\ref{scalingT205MeV}, demonstrate very good scaling behavior.

\subsection{The full profile of the electric flux tube}
In Figs.~\ref{ExNP3d_merged} the numerical results for the full profile of the nonperturbative component of the longitudinal electric field $E_x$ are displayed. In these figures we show the symmetrized version of the flux-tube profile:
\begin{equation}
    \label{symmetriprofile}
    E_x^{\rm NP}(x_l,x_t) = 
\begin{cases}
E_x^{\rm NP,\,quark}(x_l,x_t) & \text{if } x_l < d/2, \\
E_x^{\rm NP,\,antiquark}(d-x_l,x_t) & \text{if } x_l \ge d/2.
\end{cases}
\end{equation}
Remarkably, the flux-tube profile remains smooth and nearly constant (except very close to the sources) over a wide temperature range, \(43 \,\text{MeV} \lesssim T \lesssim 512 \,\text{MeV}\). However, the absolute magnitude of the electric field decreases by about two orders of magnitude when going from \(T \simeq 43\) MeV to \(T \simeq 512\) MeV.  

This behavior is illustrated in Fig.~\ref{ExNP_midpoint_vs_T}, which shows the value of \(E_x^{\mathrm{NP}}\) on the transverse midplane between the quark and antiquark sources as a function of temperature. One observes that, although the absolute value of \(E_x^{\mathrm{NP}}\) is reduced by nearly two orders of magnitude across this temperature interval, a sizable flux-tube structure persists even at the highest temperature studied.

\subsection{Field integrals: string tension and width of the flux tube}
\begin{table*}[htbp]
\begin{center} 
  \caption{Numerical results for $\sqrt{\sigma_\mathrm{eff}}$ and  $w$, as defined in Eqs.~(\ref{sigma_eff}) and~(\ref{width}).}
  \label{stringandwidth_NP}
\setlength{\tabcolsep}{12pt}
\begin{tabular}{lccrrllS[table-format=1.9]S[table-format=1.7]}
\toprule
\ \ lattice & $\beta=10/g^2$ & $a(\beta)$ [fm] & $T$ [MeV]  &  $d$ [lattice units]   & $d$ [fm] & $\sqrt{\sigma_\mathrm{eff}}$ [GeV] & $w$ [fm] \\ \midrule
$48^3 \times 48$  &  6.880  &  0.0963  &  43  \hspace{0.3cm} &  10  \hspace{0.6cm}  &  0.963  & 0.405(12)   & 0.49(4)   &   \\
$48^3 \times 16$  &  6.880  &  0.0963  & 128  \hspace{0.3cm} &  10  \hspace{0.6cm}  &  0.963  & 0.442(10)   & 0.63(5)   &   \\
$48^3 \times 14$  &  6.880  &  0.0963  & 146  \hspace{0.3cm} &  10  \hspace{0.6cm}  &  0.963  & 0.363(12)   & 0.62(8)   &   \\
$48^3 \times 12$  &  6.880  &  0.0963  & 171  \hspace{0.3cm} &   6  \hspace{0.6cm}  &  0.578  & 0.3230(11)  & 0.532(10) &   \\
$48^3 \times 12$  &  6.880  &  0.0963  & 171  \hspace{0.3cm} &   8  \hspace{0.6cm}  &  0.770  & 0.2560(31)  & 0.601(32) &   \\
$48^3 \times 12$  &  6.880  &  0.0963  & 171  \hspace{0.3cm} &  10  \hspace{0.6cm}  &  0.963  & 0.178(6)    & 0.56(9)   &   \\
$48^3 \times 12$  &  6.880  &  0.0963  & 171  \hspace{0.3cm} &  12  \hspace{0.6cm}  &  1.156  & 0.093(9)    & 0.44(12)  &   \\
$48^3 \times 12$  &  6.880  &  0.0963  & 171  \hspace{0.3cm} &  14  \hspace{0.6cm}  &  1.348  & 0.081(16)   & 0.8(4)    &   \\
$48^3 \times 10$  &  6.880  &  0.0963  & 205  \hspace{0.3cm} &  10  \hspace{0.6cm}  &  0.963  & 0.0221(8)   & 0.37(4)   &   \\
$48^3 \times  8$  &  6.880  &  0.0963  & 256  \hspace{0.3cm} &  10  \hspace{0.6cm}  &  0.963  & 0.01139(22) & 0.291(13) &   \\
$48^3 \times  6$  &  6.880  &  0.0963  & 342  \hspace{0.3cm} &  10  \hspace{0.6cm}  &  0.963  & 0.00628(9)  & 0.254(4)  &   \\
$48^3 \times  4$  &  6.880  &  0.0963  & 512  \hspace{0.3cm} &  10  \hspace{0.6cm}  &  0.963  & 0.00139(5)  & 0.160(13) &   \\
\bottomrule 
\end{tabular}
\end{center}
\end{table*}
\begin{table*}[htbp]
\begin{center} 
  \caption{Numerical results for $\sqrt{\sigma_\mathrm{eff}}$ and  $w$, as defined in Eqs.~(\ref{sigma_eff}) and~(\ref{width})  with the replacement of the nonperturbative field with the full one.}
  \label{stringandwidth}
\setlength{\tabcolsep}{12pt}
\begin{tabular}{lccrrllS[table-format=1.9]S[table-format=1.7]}
\toprule
\ \ lattice & $\beta=10/g^2$ & $a(\beta)$ [fm] & $T$ [MeV]  &  $d$ [lattice units]   & $d$ [fm] & $\sqrt{\sigma_\mathrm{eff}}$ [GeV] & $w$ [fm] \\ \midrule
$48^3 \times 48$  &  6.880  &  0.0963  &  43  \hspace{0.3cm} &  10  \hspace{0.6cm} &  0.963  & 0.517(4)     & 0.436(15)  &   \\
$48^3 \times 16$  &  6.880  &  0.0963  & 128  \hspace{0.3cm} &  10  \hspace{0.6cm} &  0.963  & 0.511(4)     & 0.45(4)    &   \\
$48^3 \times 14$  &  6.880  &  0.0963  & 146  \hspace{0.3cm} &  10  \hspace{0.6cm} &  0.963  & 0.454(4)     & 0.46(5)    &   \\
$48^3 \times 12$  &  6.880  &  0.0963  & 171 \hspace{0.3cm}  &   6  \hspace{0.6cm} &  0.578  & 0.6360(5)    & 0.418(4)   &   \\
$48^3 \times 12$  &  6.880  &  0.0963  & 171  \hspace{0.3cm} &   8  \hspace{0.6cm} &  0.770  & 0.4475(11)   & 0.412(14)  &   \\
$48^3 \times 12$  &  6.880  &  0.0963  & 171  \hspace{0.3cm} &  10  \hspace{0.6cm} &  0.963  & 0.2874(20)   & 0.446(10)  &   \\
$48^3 \times 12$  &  6.880  &  0.0963  & 171  \hspace{0.3cm} &  12  \hspace{0.6cm} &  1.156  & 0.1760(34)   & 0.468(27)  &   \\
$48^3 \times 12$  &  6.880  &  0.0963  & 171  \hspace{0.3cm} &  14  \hspace{0.6cm} &  1.348  & 0.123(5)     & 0.437(28)  &   \\
$48^3 \times 10$  &  6.880  &  0.0963  & 205  \hspace{0.3cm} &  10  \hspace{0.6cm} &  0.963  & 0.12038(30)  & 0.418(7)   &   \\
$48^3 \times  8$  &  6.880  &  0.0963  & 256  \hspace{0.3cm} &  10  \hspace{0.6cm} &  0.963  & 0.10129(9)   & 0.419(6)   &   \\
$48^3 \times  6$  &  6.880  &  0.0963  & 342  \hspace{0.3cm} &  10  \hspace{0.6cm} &  0.963  & 0.07561(4)   & 0.3807(17) &   \\
$48^3 \times  4$  &  6.880  &  0.0963  & 512  \hspace{0.3cm} &  10  \hspace{0.6cm} &  0.963  & 0.033109(19) & 0.3214(16) &   \\
\bottomrule 
\end{tabular}
\end{center}
\end{table*}
\begin{table*}[htbp]
\begin{center} 
  \caption{Numerical results for $\sqrt{\sigma_\mathrm{eff}}$ and  $w$, as defined in Eqs.~(\ref{sigma_eff}) and~(\ref{width}). Results obtained from the antiquark connected correlator.}
  \label{stringandwidth_NP_anti}
\setlength{\tabcolsep}{12pt}
\begin{tabular}{lccrrllS[table-format=1.9]S[table-format=1.7]}
\toprule
\ \ lattice & $\beta=10/g^2$ & $a(\beta)$ [fm] & $T$ [MeV]  &  $d$ [lattice units]   & $d$ [fm] & $\sqrt{\sigma_\mathrm{eff}}$ [GeV] & $w$ [fm] \\ \midrule
$48^3 \times 48$  &  6.880  &  0.0963  &  43  \hspace{0.3cm} &  10  \hspace{0.6cm}  &  0.963  & 0.374(12)   & 0.49(6)   &   \\
$48^3 \times 16$  &  6.880  &  0.0963  & 128  \hspace{0.3cm} &  10  \hspace{0.6cm}  &  0.963  & 0.384(10)   & 0.476(28) &   \\
$48^3 \times 14$  &  6.880  &  0.0963  & 146  \hspace{0.3cm} &  10  \hspace{0.6cm}  &  0.963  & 0.374(12)   & 0.65(7)   &   \\
$48^3 \times 12$  &  6.880  &  0.0963  & 171  \hspace{0.3cm} &   6  \hspace{0.6cm}  &  0.578  & 0.3139(11)  & 0.476(12) &   \\
$48^3 \times 12$  &  6.880  &  0.0963  & 171  \hspace{0.3cm} &   8  \hspace{0.6cm}  &  0.770  & 0.2529(30)  & 0.570(34) &   \\
$48^3 \times 12$  &  6.880  &  0.0963  & 171  \hspace{0.3cm} &  10  \hspace{0.6cm}  &  0.963  & 0.184(6)    & 0.54(9)   &   \\
$48^3 \times 12$  &  6.880  &  0.0963  & 171  \hspace{0.3cm} &  12  \hspace{0.6cm}  &  1.156  & 0.073(8)    & 0.283(28) &   \\
$48^3 \times 12$  &  6.880  &  0.0963  & 171  \hspace{0.3cm} &  14  \hspace{0.6cm}  &  1.348  & 0.043(26)   & 0.33(11)  &   \\
$48^3 \times 10$  &  6.880  &  0.0963  & 205  \hspace{0.3cm} &  10  \hspace{0.6cm}  &  0.963  & 0.0222(8)   & 0.59(12)  &   \\
$48^3 \times  8$  &  6.880  &  0.0963  & 256  \hspace{0.3cm} &  10  \hspace{0.6cm}  &  0.963  & 0.01365(23) & 0.45(7)   &   \\
$48^3 \times  6$  &  6.880  &  0.0963  & 342  \hspace{0.3cm} &  10  \hspace{0.6cm}  &  0.963  & 0.00628(9)  & 0.250(4)  &   \\
$48^3 \times  4$  &  6.880  &  0.0963  & 512  \hspace{0.3cm} &  10  \hspace{0.6cm}  &  0.963  & 0.00151(5)  & 0.157(12) &   \\
\bottomrule 
\end{tabular}
\end{center}
\end{table*}
\begin{table*}[htbp]
\begin{center} 
  \caption{Numerical results for $\sqrt{\sigma_\mathrm{eff}}$ and  $w$, as defined in Eqs.~(\ref{sigma_eff}) and~(\ref{width})  with the replacement of the nonperturbative field with the full one. Results obtained from the antiquark connected correlator.}
  \label{stringandwidth_anti}
\setlength{\tabcolsep}{12pt}
\begin{tabular}{lccrrllS[table-format=1.9]S[table-format=1.7]}
\toprule
\ \ lattice & $\beta=10/g^2$ & $a(\beta)$ [fm] & $T$ [MeV]  &  $d$ [lattice units]   & $d$ [fm] & $\sqrt{\sigma_\mathrm{eff}}$ [GeV] & $w$ [fm] \\ \midrule
$48^3 \times 48$  &  6.880  &  0.0963  &  43  \hspace{0.3cm} &  10  \hspace{0.6cm} &  0.963  & 0.517(4)     & 0.51(4)    &   \\
$48^3 \times 16$  &  6.880  &  0.0963  & 128  \hspace{0.3cm} &  10  \hspace{0.6cm} &  0.963  & 0.505(4)     & 0.439(18)  &   \\
$48^3 \times 14$  &  6.880  &  0.0963  & 146  \hspace{0.3cm} &  10  \hspace{0.6cm} &  0.963  & 0.418(4)     & 0.448(25)  &   \\
$48^3 \times 12$  &  6.880  &  0.0963  & 171  \hspace{0.3cm} &   6  \hspace{0.6cm} &  0.578  & 0.6361(4)    & 0.432(4)   &   \\
$48^3 \times 12$  &  6.880  &  0.0963  & 171  \hspace{0.3cm} &   8  \hspace{0.6cm} &  0.770  & 0.4482(11)   & 0.462(14)  &   \\
$48^3 \times 12$  &  6.880  &  0.0963  & 171  \hspace{0.3cm} &  10  \hspace{0.6cm} &  0.963  & 0.2875(21)   & 0.54(4)    &   \\
$48^3 \times 12$  &  6.880  &  0.0963  & 171  \hspace{0.3cm} &  12  \hspace{0.6cm} &  1.156  & 0.1631(34)   & 0.48(4)    &   \\
$48^3 \times 12$  &  6.880  &  0.0963  & 171  \hspace{0.3cm} &  14  \hspace{0.6cm} &  1.348  & 0.098(5)     & 0.43(5)    &   \\
$48^3 \times 10$  &  6.880  &  0.0963  & 205  \hspace{0.3cm} &  10  \hspace{0.6cm} &  0.963  & 0.12006(7)   & 0.436(15)  &   \\
$48^3 \times  8$  &  6.880  &  0.0963  & 256  \hspace{0.3cm} &  10  \hspace{0.6cm} &  0.963  & 0.10164(9)   & 0.430(5)   &   \\
$48^3 \times  6$  &  6.880  &  0.0963  & 342  \hspace{0.3cm} &  10  \hspace{0.6cm} &  0.963  & 0.07579(4)   & 0.3850(17) &   \\
$48^3 \times  4$  &  6.880  &  0.0963  & 512  \hspace{0.3cm} &  10  \hspace{0.6cm} &  0.963  & 0.033148(19) & 0.3190(12) &   \\
\bottomrule 
\end{tabular}
\end{center}
\end{table*}
Using the measured values of the nonperturbative electric field, and assuming cylindrical symmetry with a constant field along the flux tube, we integrate over the transverse section to obtain two quantities that quantitatively characterize the flux tube:
\begin{equation}
\begin{split}
\sigma_{\text{eff}} &= \int d^2 x_t \, \frac{(E_x^{\rm NP}(d/2,x_t))^2}{2} \\
&= \pi \int d x_t  \,\, x_t \, (E_x^{\rm NP}(d/2,x_t))^2 \,.
\end{split}
\label{sigma_eff}
\end{equation}
and
\begin{equation} 
\begin{split}
w^2 &= \frac{\int d^2 x_t \, x_t^2 \, E_x^{\rm NP}(d/2,x_t)}{\int d^2 x_t \,E_x^{\rm NP}(d/2,x_t))} \\
&= \frac{\int d x_t  \,\, x_t^3 \, E_x^{\rm NP}(d/2,x_t)}{\int d x_t  \,\, x_t \, E_x^{\rm NP}(d/2,x_t)} \,.
\end{split}
\label{width}
\end{equation}
In Eqs.~(\ref{sigma_eff}) and~(\ref{width}), the longitudinal electric field is evaluated at the midplane between the sources.  
The integral in Eq.~(\ref{sigma_eff}) represents a quantity with the dimension of energy per unit length, similar to the string tension. However, this quantity, denoted by $\sigma_{\rm eff}$, does not coincide with the true string tension, for which the integrand would involve $\sum_{a=1}^8 (E_x^a)^2$. In contrast, Eq.~(\ref{sigma_eff}) contains the squared Maxwell-like field, which is plausibly a linear combination of the Abelian color components 3 and 8 of the electric field~\cite{Baker:2024peg}.
The expression in Eq.~(\ref{width}) provides an estimate of the square of the flux-tube width.

The integrals in Eqs.~(\ref{sigma_eff}) and~(\ref{width}) are computed numerically using the trapezoidal integration algorithm. They are evaluated both for the nonperturbative component of the longitudinal electric field and for the full field. The corresponding numerical results are reported in Tables~\ref{stringandwidth_NP} and~\ref{stringandwidth}, respectively.

In the next Section, we discuss the temperature dependence of the effective string tension and of the flux-tube width.

\section{String tension, deconfinement and all that}
\label{S6}
We have collected data for QCD with (2+1)-flavors at the physical point ($m_{\pi} \; \simeq \; 140$ MeV) at finite temperatures. In our simulations we
 employed a rather large range of temperatures around the chiral pseudocritical one, $T_{\rm c} \simeq 156$ MeV, namely
 43 MeV $\lesssim  T  \lesssim $ 512 MeV, aimed to investigate the thermal behavior of the flux-tube structure between static color sources across
 the deconfinement transition. 
\begin{figure*}[htbp]
\begin{center}
\includegraphics[width=0.47\linewidth,clip]{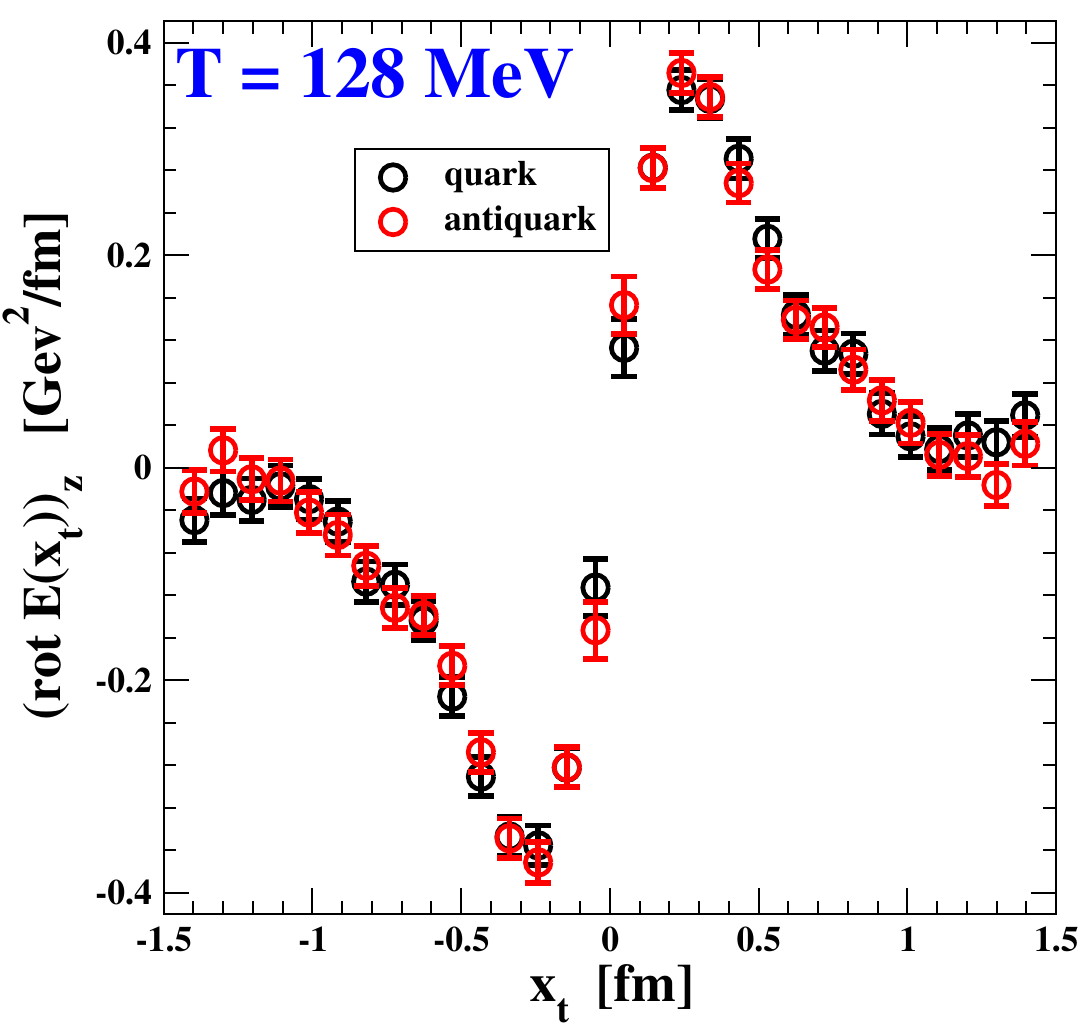}
\includegraphics[width=0.47\linewidth,clip]{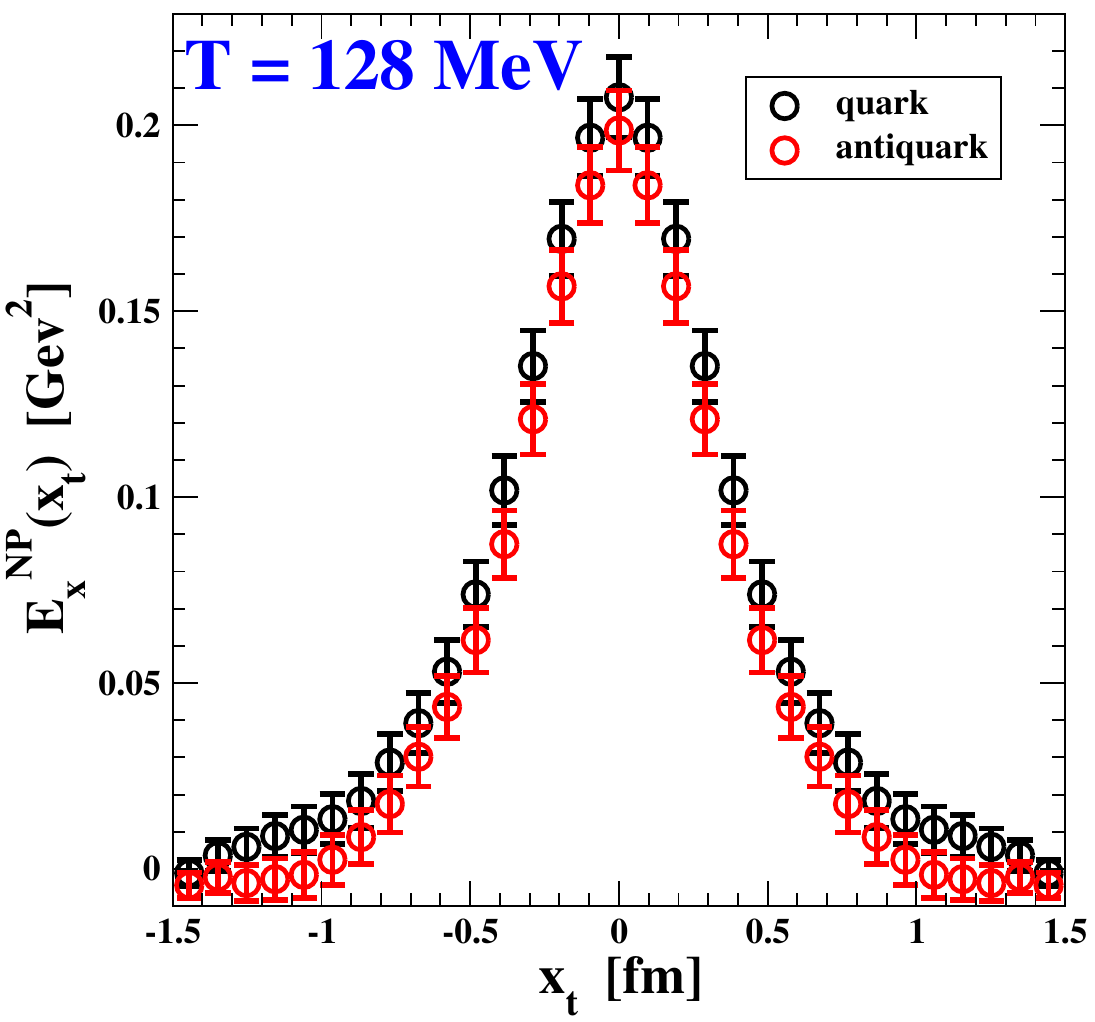}
\\
\includegraphics[width=0.47\linewidth,clip]{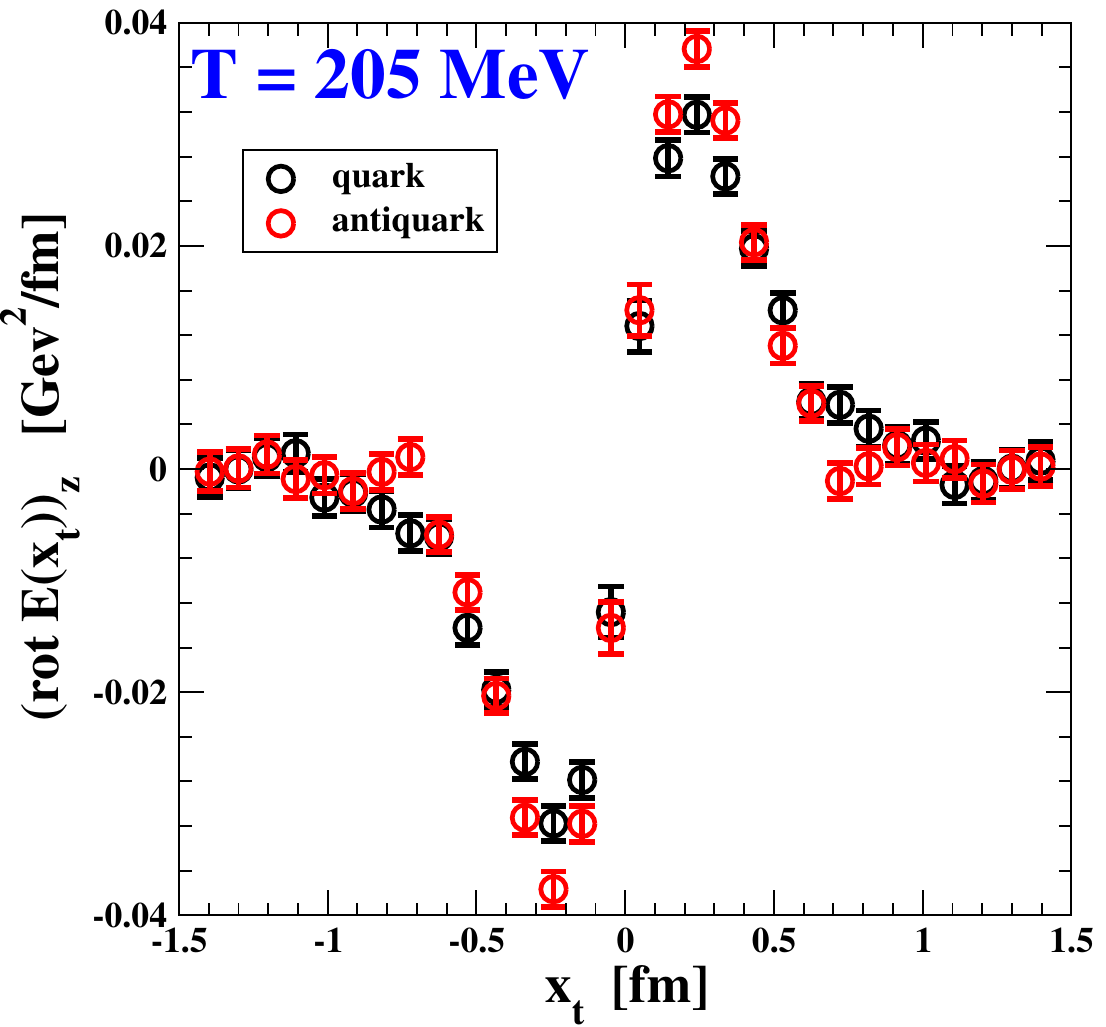}
\includegraphics[width=0.47\linewidth,clip]{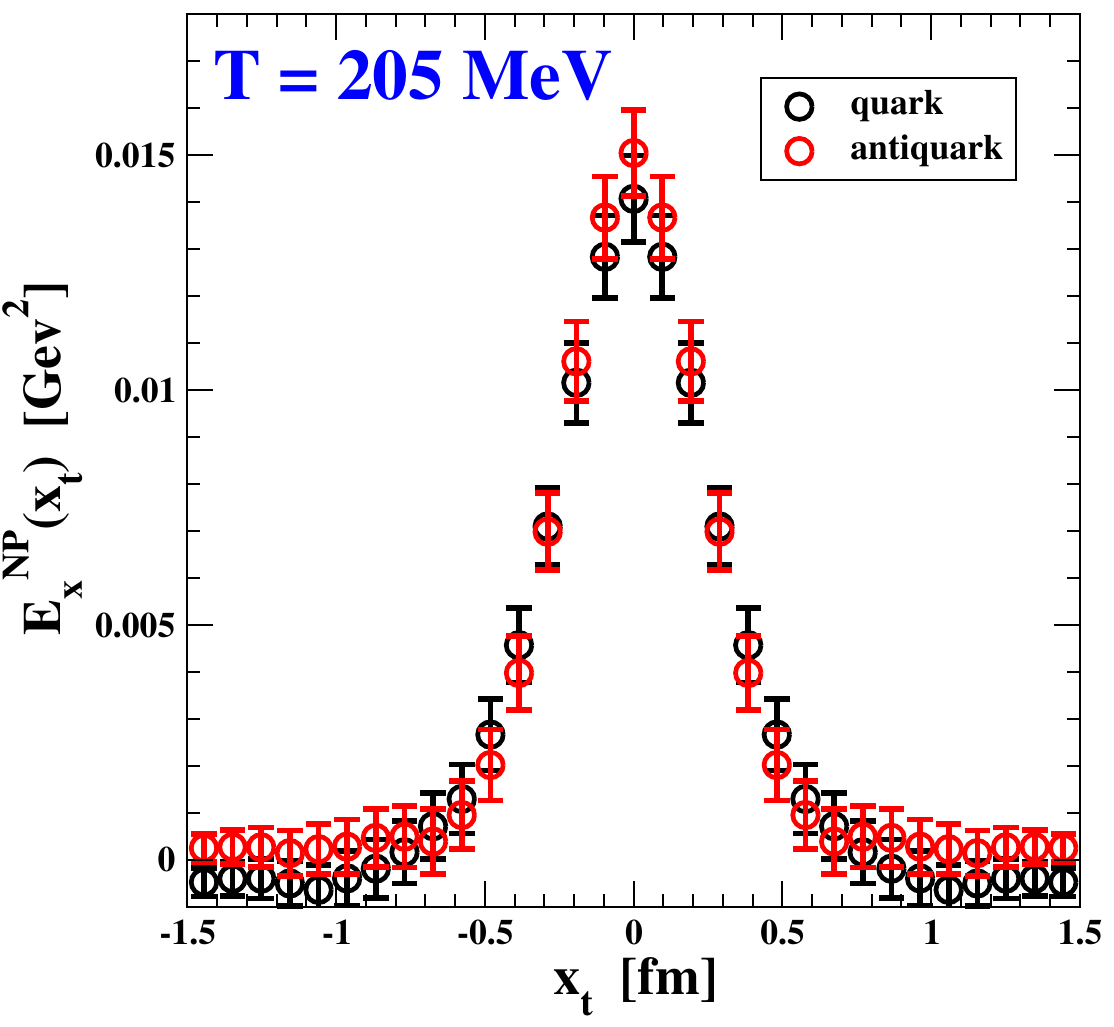}
\\
\includegraphics[width=0.47\linewidth,clip]{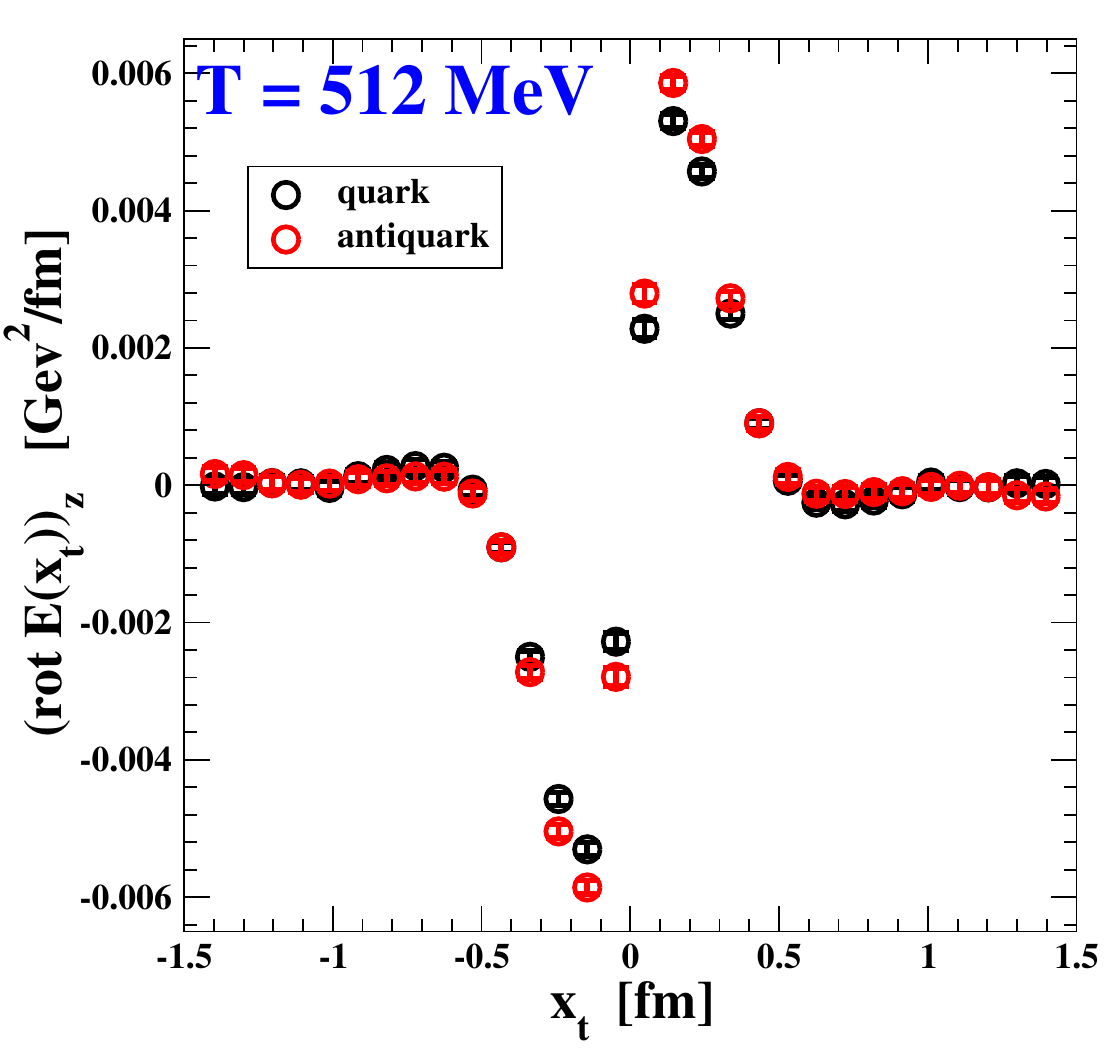}
\includegraphics[width=0.47\linewidth,clip]{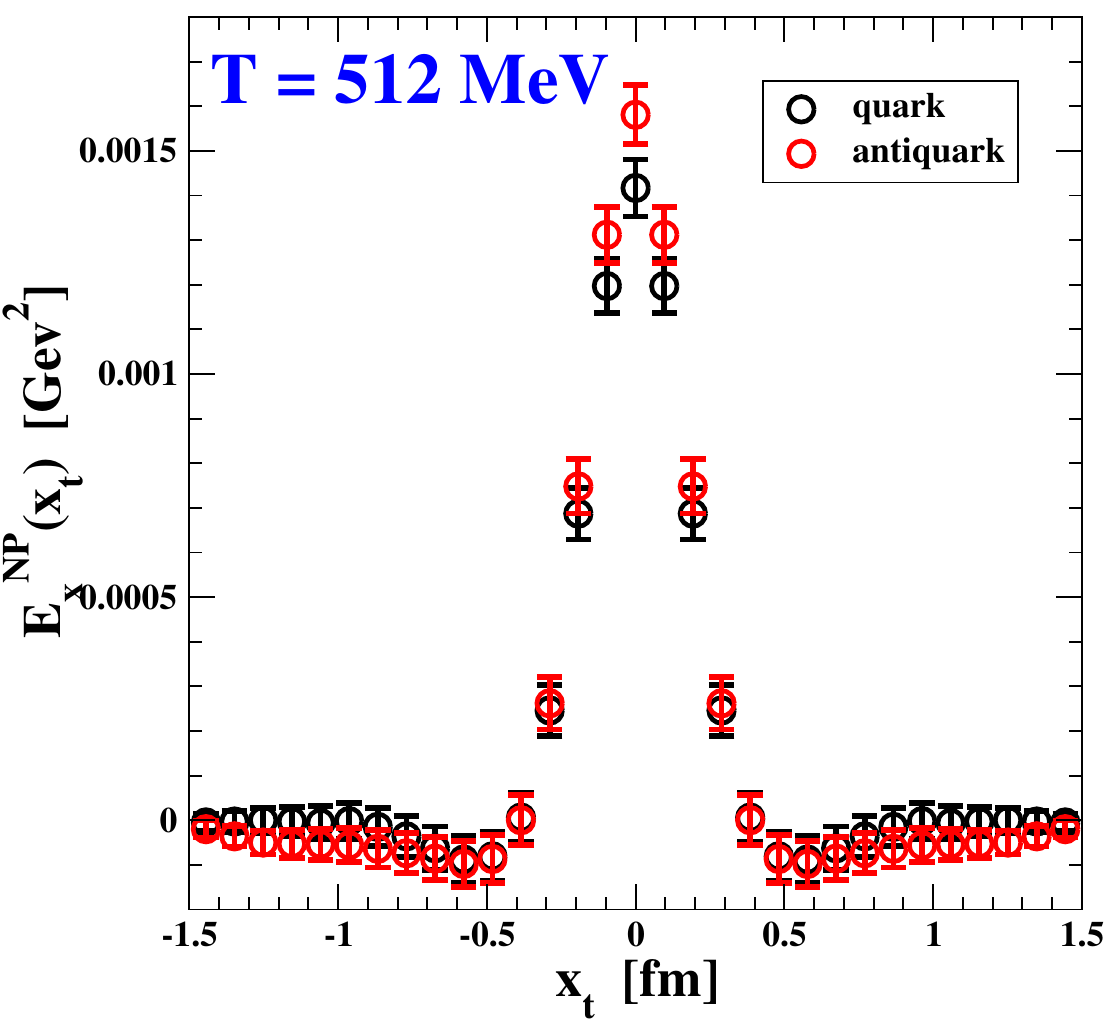}
\end{center}
\caption{Transverse distribution of the magnetic currents (left panels) and the nonperturbative electric fields (right panels) on the midplane for 
color source distance  $d \simeq 0.963$  fm  and $\beta$ = 6.880 at three different temperatures below ($T \simeq 128$ MeV) and above ($T \simeq 205, 512$ MeV) the chiral pseudocritical temperature $T_{\rm c}$. The open black circles correspond to the quark correlation function while the open red squares to the antiquark correlation
function.
}            
\label{fig:thermal}
\end{figure*}
In \cref{fig:thermal}, for illustrative purposes, we present the transverse distribution of the magnetic currents (left panels) and the nonperturbative 
electric fields (right panels) on the midplane for  color source distance  $d\simeq 0.963$  fm at three different temperatures below  and 
above the chiral pseudocritical temperature $T_{\rm c}$ obtained with both quark and antiquark connected correlation  functions.
The most surprising aspect of our results is that, even at the highest temperature employed $T \simeq 512$ MeV, there is a clear
evidence of a nonzero magnetic current and a nonperturbative longitudinal electric field that turns out to be almost uniform along
the flux-tube structure.  We have also checked that, by employing the method of Sec.~5 in Ref.~\cite{Baker:2024peg},
the nonperturbative fields are in agreement within statistical uncertainties with  \cref{1.4}, irrespectively of the values of temperatures. 
In addition, we clearly see that the shape of both the magnetic current and nonperturbative electric field does not suffer from substantial  changes
 with the temperature, unambiguously signaling the presence of the flux-tube structure. It seems that the only effect of increasing the temperature is 
a smooth decrease in the strength of magnetic currents and electric fields.
\begin{figure*}[htbp]
\begin{center}
\includegraphics[width=0.49\linewidth,clip]{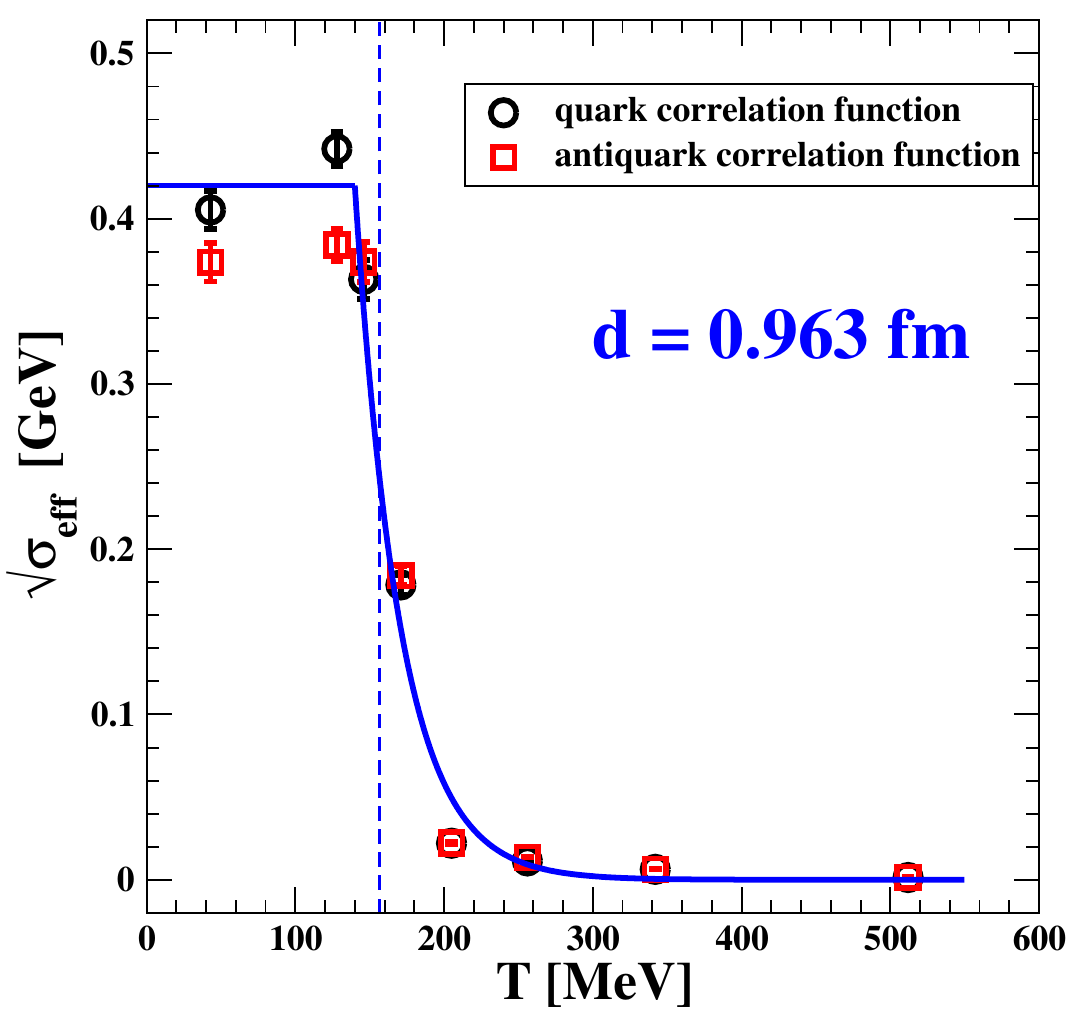}
\includegraphics[width=0.48\linewidth,clip]{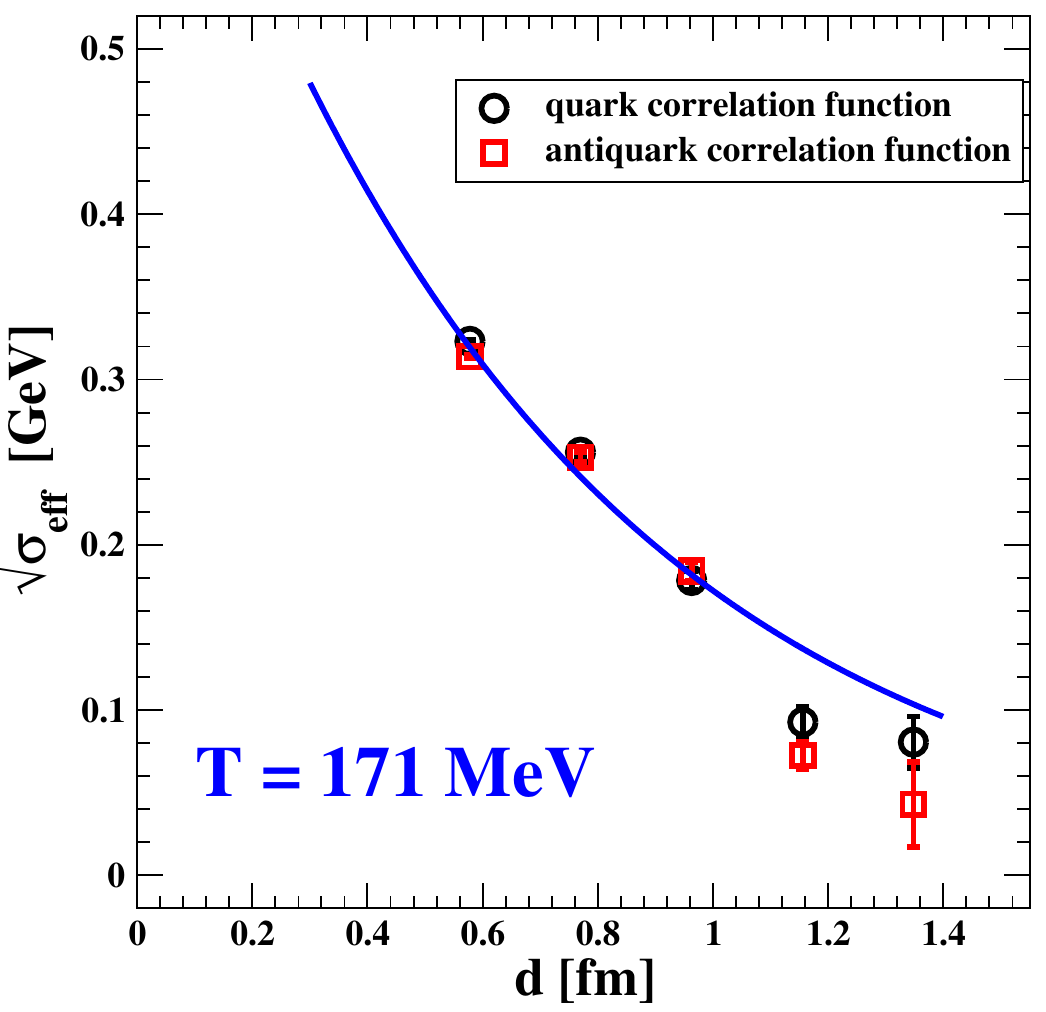}
\\
\includegraphics[width=0.49\linewidth,clip]{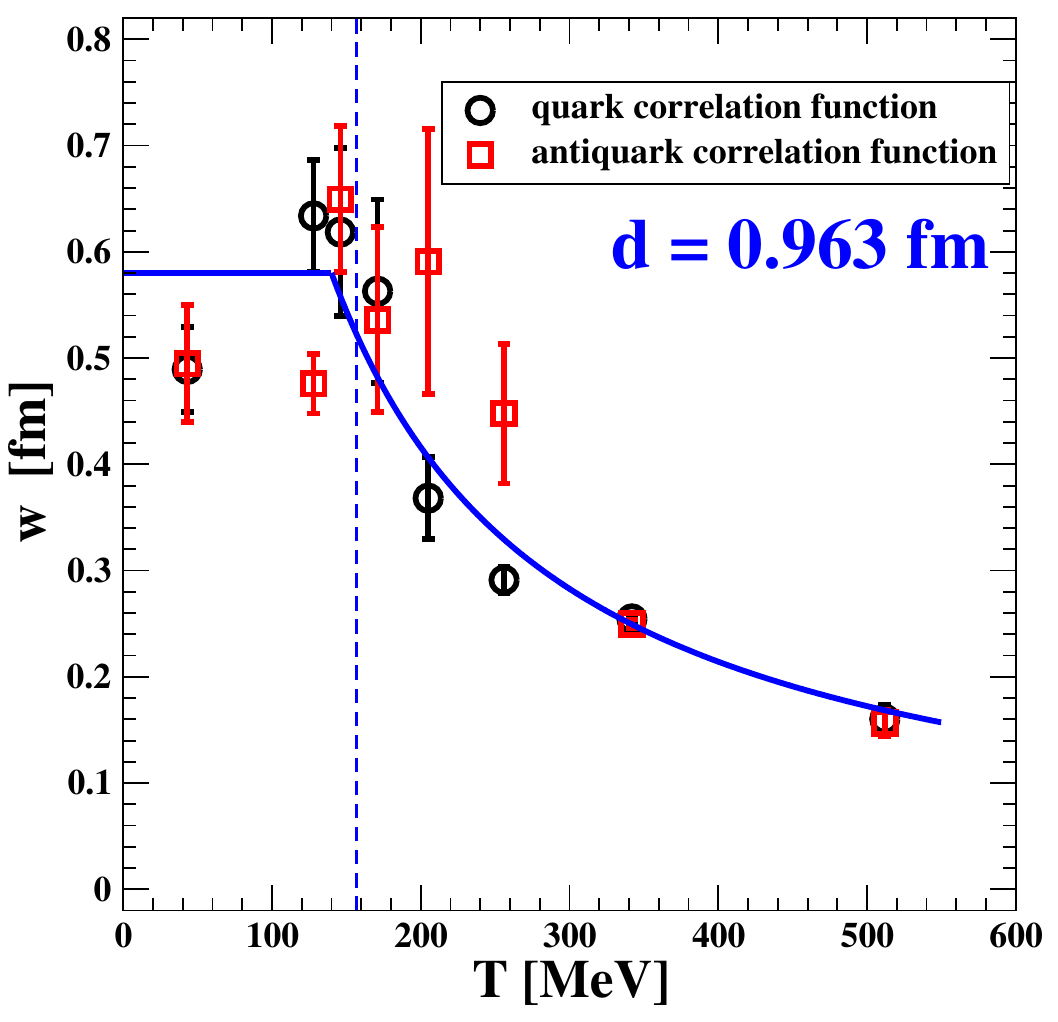}
\end{center}
\caption{(Top-left panel) The effective string tension  as given by  Eq.~(\ref{sigma_eff}) evaluated at the midplane for inter-quark distance
 $d \simeq$ 0.963 fm  at  $\beta$ = 6.880  {\it versus} the temperature. The continuous line is \cref{6.12}  with parameters given by \cref{6.15}.
(Top-right panel) Effective string tension evaluated at the midplane for $T \simeq 171$ MeV and  $\beta$ = 6.880  {\it versus} the  distance  between the static color sources. 
The continuous line is \cref{6.16} with parameters given by \cref{6.17}.
(Lower panel) The width of the flux-tube nonperturbative electric field  as given by Eq.(\ref{width})  on the midplane for inter-quark distance $d \simeq 0.963$ fm 
 and $\beta$ = 6.880 {\it versus} the temperature. The continuous line is \cref{6.18}  with parameters given by \cref{6.19}.
The open  black circles  and red squares correspond to the quark  and antiquark correlation functions, respectively.}            
\label{fig:tension}
\end{figure*}
The above  results are rather unexpected since they imply that even at temperatures well beyond the pseudocritical temperature $T_{\rm c}$ we are seeing evidence of color confinement. 
\subsection{Effective string tension}
To gain further insights, we have evaluated the effective string tension as defined by Eq.~(\ref{sigma_eff}).  In the top-left panel of \cref{fig:tension}
we report $\sqrt{\sigma_{\rm eff}}$ {\it versus} the temperature at the midplane, for quark-antiquark static sources at  $d \simeq$ 0.963 fm, 
as evaluated from the nonperturbative electric fields extracted from both the quark and antiquark connected correlation functions. 
It is evident that the thermal behavior of  $\sqrt{\sigma_{\rm eff}}$  manifests two different regimes. Below about 140 - 150 MeV the effective string tension
stays almost constant around the zero-temperature value albeit with  some scatter in the data. This means that in this low-temperature regime 
the thermal fluctuations do not modify substantially the structure of the flux-tube nonperturbative electric field and the dynamics is governed by 
wild quantum fluctuations. Note that  also in our previous paper~\cite{Baker:2024peg} in full QCD at zero temperature
we found that the effective string tension was subjected to some scatters in measured values (see Fig.~6, left panel in Ref.~\cite{Baker:2024peg}).
On the other hand, for temperatures above the pseudocritical one $T_{\rm c}$ the quantum dynamics of the hadronic system is clearly dominated
by the much smoother thermal fluctuations, such that the values of the  effective string tension extracted from the two different connected correlation functions
are in satisfying agreement.  In this high-temperature regime we see that the effective string tension manifests a drastic reduction followed by
a smoother decrease above about 200 MeV.  This kind of behavior is suggestive of an exponential decrement with the temperature.
Interestingly enough, long time ago,  the authors of Ref.~\cite{Matsui:1986dk} suggested that heavy-quarkonium suppression could be used
as  a signal of deconfinement in the quark-gluon plasma supposedly formed in high-energy heavy-ion collisions. As a matter of fact,
it was suggested that the basic mechanism for deconfinement in dense matter could originate from the Debye screening of the quark color
charge. Later on, in Ref.~\cite{Karsch:1987pv}  it was assumed that in a thermodynamic environment of interacting quarks and gluons at a given temperature $T$, quark binding becomes modified by color screening. This means, in particular, that there is a screened string tension, leading to a thermal confining
potential given by
\begin{equation}
\label{6.5}
V_{\rm conf}(R,T) \; = \; \frac{\sigma}{\mu(T)} \; \left  \{ 1 \; - \; \exp[- \mu(T) \, R] \right  \}  \;  \; ,
\end{equation}
where $\sigma$ is the zero-temperature string tension,  $\mu(T)$ is the screening mass and $R$ is the distance between the color quark-antiquark sources.
Note that, when  the screening mass tends to zero, the screened potential reduces to the familiar linear potential  $V_{\rm conf} = \sigma \, R$. On the other hand,
for $R \,  \rightarrow \, \infty$, we see that   $V_{\rm conf}(R,T)$ reduces to the temperature dependent constant $\sigma/{\mu(T)}$. \\
This kind of potential was proposed  for the first time in Ref.~\cite{Joos:1983qb}, where the form of the screened linear potential was
suggested by calculations in the Schwinger model, {\it i.e.} (1+1)-dimensional QED with massless fermions. Indeed, assuming the presence of an external dipole distribution with charge $Q$,
\begin{equation}
\label{6.6}
\rho(y) \; = \; Q \; \ \left  [  \delta \left( y - \frac{R}{2} \right)  \; - \;  \delta \left( y + \frac{R}{2} \right)  \right  ]  \;  \; ,
\end{equation}
then the potential energy is~\cite{Becher:1982ft}
\begin{equation}
\label{6.7}
V(R) \; = \; \frac{Q^2 \sqrt{\pi}}{2 \, e} \; \left  \{ 1 \; - \; \exp\left[- \frac{e}{\sqrt{\pi}} \, R\right] \right  \} \;  \; .
\end{equation}
Note that, for $e \,  \rightarrow \, 0$,
\begin{equation}
\label{6.8}
V(R) \; = \; \frac{Q^2}{2}  \; R \;  \; ,
\end{equation}
while, for $R \,  \rightarrow \, \infty$,
\begin{equation}
\label{6.9}
 V(R) \; = \; \frac{Q^2 \sqrt{\pi}}{2 \, e} \; .
\end{equation}
Equation (\ref{6.8}) is the bare Coulomb potential that in one spatial dimension is a linear confining potential with string tension 
$\sigma =  Q^2/{2 }$. The screening mass is here $\mu = e/{ \sqrt{\pi}}$. So there are no thermal effects, but the screening is due to 
quantum fluctuations. The authors of Ref.~\cite{Joos:1983qb}  used the screened potential \cref{6.5}  to reproduce the lattice data for the thermal
potential extracted from Creutz ratios in SU(2) with Wilson fermions. However, at finite temperatures large Wilson loops do not have the same interpretation as at zero temperature. So it should be necessary to replace  Wilson loops by  correlators of
Polyakov loops.  Actually, it turns out that even the more recent lattice data in full QCD obtained by employing
the Polyakov loop correlators (see, for instance, Refs.~\cite{Kaczmarek:2005ui,Borsanyi:2015yka})
seem to confirm qualitatively the behavior of the thermal potential {\it versus} the temperature reported in the quite old paper,  Ref.~\cite{Joos:1983qb}. \\
Starting from the thermal confining potential, one can define the thermal string tension as the `force' given by
\begin{equation}
\label{6.10}
\sigma(R,T) \; = \;  \frac{\partial}{\partial R } \, V_{\rm conf}(R,T)  \;  \; .
\end{equation}
Using  \cref{6.5}, we get
\begin{equation}
\label{6.11}
\sigma(R,T) \; = \;  \sigma \;  \exp[- \mu(T) \, R]   \;  \; .
\end{equation}
Evidently, since at zero temperature there is no screening, $\sigma$ is the zero-temperature screening  string tension. 
We see, thus, that our lattice data could be interpreted as evidence of a  string tension  screened by thermal fluctuations 
with a suitable screening mass $\mu(T)$.\\
As we already pointed out,  at high temperatures and zero baryon-number density the hadronic matter is expected to undergo a phase change
to a new form of matter, the quark-gluon plasma. As is well known, the nature of the degree of freedom composing the thermal hadronic system
at high temperatures is encoded in the spectral functions (see, for instance, the recent review Ref.~\cite{Bazavov:2020teh} and references therein). In fact, since at high temperature it is expected that quarks and gluons  deconfine, this 
circumstance is  intimately reflected in properties of the hadron correlation functions and the thermal masses extracted from them.
Lattice measurements in QCD with dynamical fermions of spatial correlations and related hadronic screening masses  
have been performed for the first time in Refs.~\cite{Detar:1987kae,Detar:1987hib}.  More recently, the authors of Ref.~\cite{Bazavov:2019www}
presented  lattice results for  mesonic screening masses in the temperature range 140 MeV  $\lesssim T  \lesssim$ 2500 MeV 
for QCD with (2+1)-flavor of dynamical  fermions at the physical point. In particular, looking at Fig.~5 in Ref.~\cite{Bazavov:2019www}, it seems to us that
the low-lying mesonic thermal screening masses manifest themselves above a certain onset temperature smaller but very close to the chiral
critical temperature $T_{\rm c}$, and after that there is an almost linear  increase  with the temperature.
This looks quite similar to the behavior of the thermal effective string tension that, as pointed out before, stays
almost constant to the zero-temperature value for temperatures below about 140 - 150 MeV.  As a consequence, according to \cref{6.11}, we are led
to assume that
\begin{equation}
\label{6.12}
\sqrt{\sigma_{\rm eff}}(d,T) \; = \;  \sqrt{\sigma_{\rm eff}}(0) \;  \exp[- \frac{1}{2} \, \mu_{\rm st}(T) \, d]   \;  \; ,
\end{equation}
where $d$ is the distance between the static color sources,   $\sqrt{\sigma_{\rm eff}}(0)$ the zero-temperature effective string tension and
\begin{equation}
\label{6.13}
 \mu_{\rm st}(T) \; \simeq \;  
 \left\{ \begin{array}{ll}
 \; \;   0  \; \; &   \; \; T  \; \lesssim \; T_0  \; ,
  \\
 \;  \; a_{\rm st} \; (T \; - \; T_0)  \; \; &  \; \; T_0  \; \lesssim \; T\;.
\end{array}
    \right.
\end{equation}
We choose to fix the onset temperature $T_0$ to the QCD mass gap:
\begin{equation}
\label{6.14}
T_0  \; \simeq \; m_{\pi}  \;  \simeq \; 140 \; \text{MeV} \; \; .
\end{equation}
After that, the fit of the lattice data for $\sqrt{\sigma_{\rm eff}}$ to \cref{6.12} returned
\begin{equation}
\label{6.15}
 \sqrt{\sigma_{\rm eff}}(0) \;  \simeq \; 0.42 \; \text{GeV} \; \; , \; \;  a_{\rm st} \;  = \; 13.5(1) \; \; .
\end{equation}
Indeed, \cref{6.12} seems to track quite well the lattice data (see the continuous line in Fig.~\ref{fig:tension}, top-left panel)
albeit with a rather high reduced chi-square ($\chi^2_r \sim 10^3$). \\
We have also performed a further check of \cref{6.12} by fixing the temperature to $T_1 \, \simeq \, 171$ MeV and measuring the
effective string tension for various color source distances as displayed in  Fig.~\ref{fig:tension}, top-right panel.
The lattice data have been fitted to
\begin{equation}
\label{6.16}
\sqrt{\sigma_{\rm eff}}(d,T_1) \; = \;  \alpha_{\rm st}  \;  \exp[- \frac{1}{2} \, \mu_{\rm st}(T_1) \, d]   \;  \; .
\end{equation}
From the best fit we obtained
%
\begin{equation}
\label{6.17}
 \alpha_{\rm st}  \;  =  \; 0.743(28) \; \text{GeV} \; \; , \; \;  a_{\rm st} \;  = \; 18.6(8) \; \; 
\end{equation}
with reduced chi-square  $\chi^2_r \sim 10$. The parameter  $ \alpha_{\rm st}$ is devoid of physical significance since it would correspond to
the value of the effective string tension extrapolated to zero color source distance. In addition,  the best-fitted parameter
 $a_{\rm st}$ is somewhat higher than  the previous estimate in \cref{6.15}. It is worthwhile to note that   $ a_{\rm st}$ measures the
 ratio of the screening mass over the temperature for $T  \gg T_0$.  It is, also, interesting to mention that our first determination for this ratio, 
 given by \cref{6.15}, seems  to  be compatible with the ratio $m_E/T = 13.0(1.1)$ between the electric screening mass and the temperature as reported in 
 Ref.~\cite{Maezawa:2010vj}  in the case of lattice QCD with two flavors of degenerate Wilson fermions,  
 but it is substantial different from $m_E/T  = 7.31(25)$ reported in  Ref.~\cite{Borsanyi:2015yka}  for  lattice QCD with (2+1)-flavor of  
 stout improved staggered quarks at the physical point  after continuum extrapolation. 
\subsection{Flux-tube width}
The almost exponential decrease of the effective string tension above the chiral pseudocritical temperature is not the whole story for the thermal
behavior of the flux tube generated by two static color sources.  Actually, we have also looked  at the thermal behavior of the flux-tube width
as defined by \cref{width}. In Fig.~\ref{fig:tension}, lower panel, we display  the width evaluated at  the midplane for a fixed static source  distance $d \simeq $0.963 fm 
{\it versus} the temperature.  As in the case of the effective string tension, in the low-temperature region there are sizable fluctuations, similarly to what
observed in our previous paper (compare with Fig.~6, right  panel in Ref.~\cite{Baker:2024peg}), while well above $T_{\rm c}$
lattice data are more stable.  The most noticeable aspect of our data is that the flux-tube width decreases with the temperature
in the putative deconfined region. This is in striking contrast with standard lore where  the width of the flux tube should increase with the temperature
until, near the deconfinement temperature, it should  become comparable to the distance between the static color sources, washing out the
whole flux-tube structure. \\
We have, also, attempted a mere  phenomenological fit to the lattice data according to
\begin{equation}
\label{6.18}
 w(T) \; \simeq \;  
 \left\{ \begin{array}{ll}
 \; \;   w(0)  \; \; &   \; \; T  \; \lesssim \; T_0  \;,
  \\
 \;  \; \frac{w(0)}{1 \; + \; a_{\rm w}  \; (T \; - \; T_0)}  \; \; &  \; \; T_0  \; \lesssim \; T \;,
\end{array}
    \right.
\end{equation}
obtaining
\begin{equation}
\label{6.19}
w(0) \;  \simeq  \; 0.58  \; \;  \text{fm} \; \; , \; \;  a_{\rm w} \;  = \; 0.00657(25)  \; \;  \text{MeV}^{-1} \;  \;,
\end{equation}
again with a rather sizable reduced chi-square, $\chi^2_r \sim 10$. \\

In a recent paper~\cite{Galsandorj:2023rqw} flux tubes were studied in (2 + 1)-flavor QCD with physical mass at high temperature. Differently from us, the gradient flow method was used to increase the signal-to-noise ratio and the characterization of flux tubes was based on the full longitudinal electric field ({\it i.e.} without the subtraction of the perturbative component). Moreover, the highest temperature considered was about 200 MeV. In that range of temperatures, their results for the width of the flux tube are comparable with ours.

\section{Conclusions}
Lattice QCD indicates that, at temperatures around and above the hadronic scale, strongly interacting matter undergoes a \emph{crossover} near \(T_{\rm c} \simeq 156\,\mathrm{MeV}\) from hadronic to quark matter, with deconfinement and chiral restoration reflected in a rapid rise of thermodynamic observables and peaks in quark-number susceptibilities. Polyakov-loop correlators show Debye screening well above \(T_{\rm c}\)~\cite{Borsanyi:2015yka}, though bulk quantities still deviate from Stefan--Boltzmann limits, and heavy-ion data reveal a nearly perfect fluid rather than a weakly interacting gas (see, {\it e.g.},~\cite{Heinz:2008tv}).

In the present paper, we studied the thermal behavior of the flux-tube structures generated by a static quark-antiquark pair separated by
distances 0.57 fm $\lesssim  d \lesssim$  1.348 fm. We varied the temperature in the rather ample interval 
43  MeV $\lesssim  T \lesssim$  512 MeV that includes the chiral pseudocritical temperature $T_{\rm c} \simeq $ 156 MeV.

Our main results are summarized as follows:
\begin{itemize}
\item[$\bullet$] We observed the fingerprint of squeezed flux-tube for all the employed temperatures, as given by the presence of
magnetic currents, and a longitudinal nonperturbative electric field almost uniform along the flux tube. The main effects observed by increasing
the temperature above the chiral pseudocritical temperature are the further squeezing of the flux-tube transverse distribution and the monotonic
decrease in the strengths of both the magnetic current and the nonperturbative electric field.
\vspace{0.1cm}
\item[$\bullet$]  We found  that the effective string tension and the flux-tube width stay almost constant at their zero-temperature
values up to a certain onset temperature of about 140 - 150 MeV. After that, the string tension exhibits an almost exponential suppression with increasing temperatures, while the flux-tube width decreases following an almost inverse-temperature
law.
\vspace{0.1cm}
\item[$\bullet$]  For temperatures well above the chiral pseudocritical temperature the thermal hadronic system can be thought as an ensemble of 
elementary degrees of freedom constrained to colorless bound states characterized by tiny and feeble flux-tube structures.
\end{itemize}

According to our results  the notion of a fully ``free'' deconfined vacuum should be reconsidered, since we find  narrow electric flux tubes between static \(Q\bar Q\) sources persisting up to \(T \simeq 512\,\mathrm{MeV}\). Above an onset temperature \(T_0 \lesssim T_{\rm c}\), a \emph{screened confining phase} emerges in which the effective string tension falls almost exponentially with \(T\), while the tube width scales as \(\sim 1/T\). This suggests that confining mechanisms remain active across \(T_{\rm c}\), {\it i.e.}, there is no sharp vacuum change.

Phenomenologically, for 200 MeV $\lesssim T \lesssim$ 500 MeV we estimate \(\sqrt{\sigma_{\rm eff}} \lesssim 10^{-2}\,\mathrm{GeV}\), implying color-singlet states of size larger than 10 fm: the partition function can be viewed in terms of widely separated partons bound into singlets, yielding a strongly interacting fluid, consistent with RHIC/LHC observations. Extrapolating our trend, a weakly interacting quark--gluon plasma would emerge only at much higher temperatures, well above \(\sim 1\,\mathrm{GeV}\).

\section*{Acknowledgments}
This investigation was in part based on the MILC collaboration's public lattice gauge theory code (\url{https://github.com/milc-qcd/}). Numerical calculations have been made possible through a CINECA-INFN agreement, providing access to HPC resources at CINECA. PC, LC and AP acknowledge support from INFN/NPQCD project. VC acknowledges support by  the Deutsche Forschungsgemeinschaft \linebreak (DFG, German Research Foundation) through the CRC-TR 211  ``Strong-interaction matter under extreme conditions'' -- \linebreak project number 315477589 -- TRR 211. This work is (partially) supported by ICSC – Centro Nazionale di Ricerca in High Performance Computing, Big Data and Quantum Computing, funded by European Union – NextGenerationEU.

\bibliography{qcd}

\begin{thebibliography}{41}

\bibitem{Baker:2018mhw}
M.~Baker, P.~Cea, V.~Chelnokov, L.~Cosmai, F.~Cuteri, A.~Papa, Eur. Phys. J.
  \textbf{C79}, 478 (2019), \texttt{1810.07133}

\bibitem{Baker:2019gsi}
M.~Baker, P.~Cea, V.~Chelnokov, L.~Cosmai, F.~Cuteri, A.~Papa, Eur. Phys. J. C
  \textbf{80}, 514 (2020), \texttt{1912.04739}

\bibitem{Baker:2022cwb}
M.~Baker, V.~Chelnokov, L.~Cosmai, F.~Cuteri, A.~Papa, Eur. Phys. J. C
  \textbf{82}, 937 (2022), \texttt{2207.08797}

\bibitem{Baker:2024peg}
M.~Baker, P.~Cea, V.~Chelnokov, L.~Cosmai, A.~Papa, Eur. Phys. J. C
  \textbf{85}, 29 (2025), \texttt{2409.20168}

\bibitem{Baker:2023dnn}
M.~Baker, V.~Chelnokov, L.~Cosmai, F.~Cuteri, A.~Papa, Eur. Phys. J. C
  \textbf{84}, 150 (2024), \texttt{2310.04298}

\bibitem{Bierlich:2022oja}
C.~Bierlich, S.~Chakraborty, G.~Gustafson, L.~L\"onnblad, SciPost Phys.
  \textbf{13}, 023 (2022), \texttt{2202.12783}

\bibitem{Bierlich:2020naj}
C.~Bierlich, S.~Chakraborty, G.~Gustafson, L.~L\"onnblad, JHEP \textbf{03}, 270
  (2021), \texttt{2010.07595}

\bibitem{Cea:2023}
P.~{Cea}, Universe \textbf{2024}, 10, 111 (2024), \texttt{2311.14791}

\bibitem{Bazavov:2017dus}
A.~Bazavov et~al., Phys. Rev. D \textbf{95}, 054504 (2017), \texttt{1701.04325}

\bibitem{Aoki:2006we}
Y.~Aoki, G.~Endrodi, Z.~Fodor, S.D. Katz, K.K. Szabo, Nature \textbf{443}, 675
  (2006), \texttt{hep-lat/0611014}

\bibitem{Bhattacharya:2014ara}
T.~Bhattacharya et~al., Phys. Rev. Lett. \textbf{113}, 082001 (2014),
  \texttt{1402.5175}

\bibitem{HotQCD:2018pds}
A.~Bazavov et~al. (HotQCD), Phys. Lett. B \textbf{795}, 15 (2019),
  \texttt{1812.08235}

\bibitem{Aoki:2006br}
Y.~Aoki, Z.~Fodor, S.D. Katz, K.K. Szabo, Phys. Lett. B \textbf{643}, 46
  (2006), \texttt{hep-lat/0609068}

\bibitem{Gross:2022hyw}
F.~Gross et~al., Eur. Phys. J. C \textbf{83}, 1125 (2023), \texttt{2212.11107}

\bibitem{Aarts:2023vsf}
G.~Aarts et~al., Prog. Part. Nucl. Phys. \textbf{133}, 104070 (2023),
  \texttt{2301.04382}

\bibitem{DiGiacomo:1989yp}
A.~Di~Giacomo, M.~Maggiore, S.~Olejnik, Phys. Lett. \textbf{B236}, 199 (1990)

\bibitem{Skala:1996ar}
P.~Skala, M.~Faber, M.~Zach, Nucl.Phys. \textbf{B494}, 293 (1997),
  \texttt{hep-lat/9603009}

\bibitem{Cea:2013oba}
P.~Cea, L.~Cosmai, F.~Cuteri, A.~Papa, PoS \textbf{LATTICE2013}, 468 (2013),
  \texttt{1310.8423}

\bibitem{Cea:2014uja}
P.~Cea, L.~Cosmai, F.~Cuteri, A.~Papa, Phys. Rev. \textbf{D89}, 094505 (2014),
  \texttt{1404.1172}

\bibitem{Cea:2014hma}
P.~Cea, L.~Cosmai, F.~Cuteri, A.~Papa, PoS \textbf{LATTICE2014}, 350 (2014),
  \texttt{1410.4394}

\bibitem{Cea:2015wjd}
P.~Cea, L.~Cosmai, F.~Cuteri, A.~Papa, JHEP \textbf{06}, 033 (2016),
  \texttt{1511.01783}

\bibitem{Jahn:2004qr}
O.~Jahn, O.~Philipsen, Phys. Rev. D \textbf{70}, 074504 (2004),
  \texttt{hep-lat/0407042}

\bibitem{Follana:2006rc}
E.~Follana et~al. (HPQCD Collaboration, UKQCD Collaboration), Phys.Rev.
  \textbf{D75}, 054502 (2007), \texttt{hep-lat/0610092}

\bibitem{Bazavov:2009bb}
A.~Bazavov et~al. (MILC), Rev. Mod. Phys. \textbf{82}, 1349 (2010),
  \texttt{0903.3598}

\bibitem{Bazavov:2010ru}
A.~Bazavov et~al. (MILC), Phys. Rev. \textbf{D82}, 074501 (2010),
  \texttt{1004.0342}

\bibitem{Bazavov:2011nk}
A.~Bazavov et~al., Phys. Rev. \textbf{D85}, 054503 (2012), \texttt{1111.1710}

\bibitem{MILC:2010hzw}
A.~Bazavov et~al. (MILC), PoS \textbf{LATTICE2010}, 074 (2010),
  \texttt{1012.0868}

\bibitem{Hasenfratz:2001hp}
A.~Hasenfratz, F.~Knechtli, Phys. Rev. \textbf{D64}, 034504 (2001),
  \texttt{hep-lat/0103029}

\bibitem{Matsui:1986dk}
T.~Matsui, H.~Satz, Phys. Lett. B \textbf{178}, 416 (1986)

\bibitem{Karsch:1987pv}
F.~Karsch, M.T. Mehr, H.~Satz, Z. Phys. C \textbf{37}, 617 (1988)

\bibitem{Joos:1983qb}
H.~Joos, I.~Montvay, Nucl. Phys. B \textbf{225}, 565 (1983)

\bibitem{Becher:1982ft}
P.~Becher, Annals Phys. \textbf{146}, 223 (1983)

\bibitem{Kaczmarek:2005ui}
O.~Kaczmarek, F.~Zantow, Phys. Rev. \textbf{D71}, 114510 (2005),
  \texttt{hep-lat/0503017}

\bibitem{Borsanyi:2015yka}
S.~Bors{\'a}nyi, Z.~Fodor, S.D. Katz, A.~P{\'a}sztor, K.K. Szab{\'o},
  C.~T{\"o}r{\"o}k, JHEP \textbf{04}, 138 (2015), \texttt{1501.02173}

\bibitem{Bazavov:2020teh}
A.~Bazavov, J.H. Weber, Prog. Part. Nucl. Phys. \textbf{116}, 103823 (2021),
  \texttt{2010.01873}

\bibitem{Detar:1987kae}
C.E. Detar, J.B. Kogut, Phys. Rev. Lett. \textbf{59}, 399 (1987)

\bibitem{Detar:1987hib}
C.E. Detar, J.B. Kogut, Phys. Rev. D \textbf{36}, 2828 (1987)

\bibitem{Bazavov:2019www}
A.~Bazavov et~al., Phys. Rev. D \textbf{100}, 094510 (2019),
  \texttt{1908.09552}

\bibitem{Maezawa:2010vj}
Y.~Maezawa, S.~Aoki, S.~Ejiri, T.~Hatsuda, N.~Ishii, K.~Kanaya, N.~Ukita,
  T.~Umeda (WHOT-QCD), Phys. Rev. D \textbf{81}, 091501 (2010),
  \texttt{1003.1361}

\bibitem{Galsandorj:2023rqw}
E.~Galsandorj, S.~Chagdaa, B.~Purev, Phys. Part. Nucl. Lett. \textbf{20}, 10
  (2023)

\bibitem{Heinz:2008tv}
U.W. Heinz, J. Phys. A \textbf{42}, 214003 (2009), \texttt{0810.5529}

\end{thebibliography}

\end{document}